\documentclass[apj]{emulateapj}
\usepackage{natbib}
\newcommand{\av}{A$_{\textnormal{v}}$}

\begin{document}

\title{The Spitzer c2d Survey of Large, Nearby, Interstellar Clouds. XII. 
The Perseus YSO Population as Observed with IRAC and MIPS}

\author{Kaisa E. Young and Chadwick H. Young}
\affil{Department of Physical Sciences, Nicholls State University, P.O. Box 2022, Thibodaux, LA 70310, USA}
\email{kaisa.young@nicholls.edu}
\email{chad.young@nicholls.edu}
\author{Shih-Ping Lai}
\affil{Institute of Astronomy and Department of Physics, National Tsing Hua University, Hsinchu 30013, Taiwan}
\email{slai@phys.nthu.edu.tw}
\author{Michael M. Dunham}
\affil{Harvard-Smithsonian Center for Astrophysics, 60 Garden Street, MS 78, Cambridge, MA 02138, USA}
\email{mdunham@cfa.harvard.edu}
\and
\author{Neal J. Evans II}
\affil{Department of Astronomy, University of Texas at Austin, 2515 Speedway, Stop C1400, Austin, TX 78712, USA}
\email{nje@astro.as.utexas.edu}

\begin{abstract}
\textsc{}
The {\it Spitzer Space Telescope} mapped the Perseus molecular cloud complex with IRAC and MIPS as part of the c2d {\it Spitzer} Legacy project. This paper combines the observations from both instruments giving an overview of low-mass star formation across Perseus from 3.6 to 70 \micron. We provide an updated list of young stellar objects with new classifications and source fluxes from previous works, identifying 369 YSOs in Perseus with the {\it Spitzer} dataset.
By synthesizing the IRAC and MIPS maps of Perseus and building on the work of previous papers in this series \citep{jor06,reb07}, we present a current census of star formation across the cloud and within smaller regions.  67\% of the YSOs are associated with the young clusters NGC 1333 and IC 348. The majority of the star formation activity in Perseus occurs in the regions around the clusters, to the eastern and western ends of the cloud complex. The middle of the cloud is nearly empty of YSOs despite containing regions of high visual extinction. The western half of Perseus contains three-quarters of the total number of embedded YSOs (Class 0+I and Flat SED sources) in the cloud and nearly as many embedded YSOs as Class II and III sources. Class II and III greatly outnumber Class 0+I objects in eastern Perseus and IC 348. These results are consistent with previous age estimates for the clusters. Across the cloud, 56\% of YSOs and 91\% of the Class 0+I and Flat sources are in areas where \av\ $\geq$ 5 mag, indicating a possible extinction threshold for star formation.
\end{abstract}


\section{Introduction}

Large-scale studies of nearby molecular clouds provide the ability to study the environment of the earliest stages of star formation. The ``From Molecular Cores to Planet Forming Disks'' (c2d) \textit{Spitzer} Legacy project \citep{eva03} used the \textit{Spitzer Space Telescope} \citep{wer04} to map five large molecular clouds with the Infrared Array Camera (IRAC; \citealt{faz04}) and the Multi-Band Imaging Photometer for \textit{Spitzer} (MIPS; \citealt{rie04}) including Serpens, Chamaeleon II, Lupus, Ophiuchus, and Perseus. The c2d team's data and results for all five large clouds were compared and summarized in \citet{eva09}, hereafter E09.

We present a study of the complete IRAC and MIPS c2d maps of the Perseus molecular cloud. Our work synthesizes the data sets first presented by \citet{jor06} and \citet{reb07} putting the full \textit{Spitzer} observations in context of more recent work on Perseus. This paper gives an overview of low-mass star formation across the varied environments in Perseus from 3.6 to 70 \micron. We provide an updated list of young stellar objects (YSOs) with newly corrected photometry, spectral indices, and classifications.

Perseus is a well-known and a much-studied cloud \citep[e.g.,][]{sar79, lad93, kir06, eno07} because it presents a range of star formation environments. Perseus contains two young clusters, IC 348 and NGC 1333, as well as several small dense clumps of the type that often produce one or a few stars (B1, B5, L1448, and L1455). Perseus is near the middle of the spectrum of star formation in large clouds that spans from a low density (e.g., Chamaeleon II) to a high density (e.g., Serpens) of young stellar objects (\citealt{eva09}). Perseus is not currently forming very high-mass (mid-B or earlier spectral type) stars; although, IC 348 is known to contain a B5 V star (BD+31\degr 634). Therefore, Perseus has been called a ``typical'' cloud in the study of the early phases of low-mass star formation \citep{eno06}.

Perseus also contains protostellar cores, very low luminosity objects (VeLLOs), and potentially very young YSOs that are too deeply embedded to have been detected by {\it Spitzer}. Six candidate first hydrostatic cores have been identified in Perseus \citep[e.g.,][]{che10, eno10, pin11}. \citet{dun08} identify numerous low-luminosity embedded protostars and as well as VeLLOs in the cloud. Enoch et al. \citeyearpar{eno06} combined the \textit{Spitzer} map with a 1.1 mm Bolocam dust continuum map of Perseus from the {\it Caltech Submillimeter Observatory} (CSO). They discuss the properties of prestellar \citep{eno08} and the Class 0 and I cores \citep{eno09} identified with the \textit{Spitzer} and Bolocam data. Perseus was also mapped in the submillimeter using the Submillimeter Common User Bolometer Array (SCUBA) at the {\it James Clerk Maxwell Telescope} \citep[JCMT;][]{hat05,kir06}. Discussion of the \textit{Spitzer} sources in relation to the SCUBA cores can be found in \citet{hat07} and \citet{jor07}. More recently, \citet{sad14} identified 28 candidate Class 0 sources in Perseus using the {\it Herschel Space Observatory}. Four of these sources had not been previously identified with \textit{Spitzer}. 

The IRAC and MIPS data sets for Perseus were analyzed individually by \citet{jor06} and \citet{reb07} respectively. This is the 12th paper in a series presenting original \textit{Spitzer} data of the five large molecular clouds mapped by the c2d team. We adopt the same distance of 250 $\pm$ 50 pc for Perseus \citep{eno06} for consistency with previous work by the c2d team. However, very long baseline interferometry (VLBI) studies of H$_2$O masers in nearby star-forming regions give a slightly closer distance to the western end of Perseus: 235 $\pm$ 18 pc for NGC 1333 \citep{hir08} and 232 $\pm$ 18 pc to L1448 \citep{hir11}. 

We follow the same data reduction and analysis procedures laid out in previous papers in the series and briefly described in the following sections. We have updated the c2d list of YSOs in Perseus, providing new extinction corrected photometry, and discuss the classification and distribution of YSOs both within and outside the clusters in the context of the extended emission and visual extinction in Perseus.

\section{Observations and Data}

\begin{figure}
\figurenum{1}
\plotone{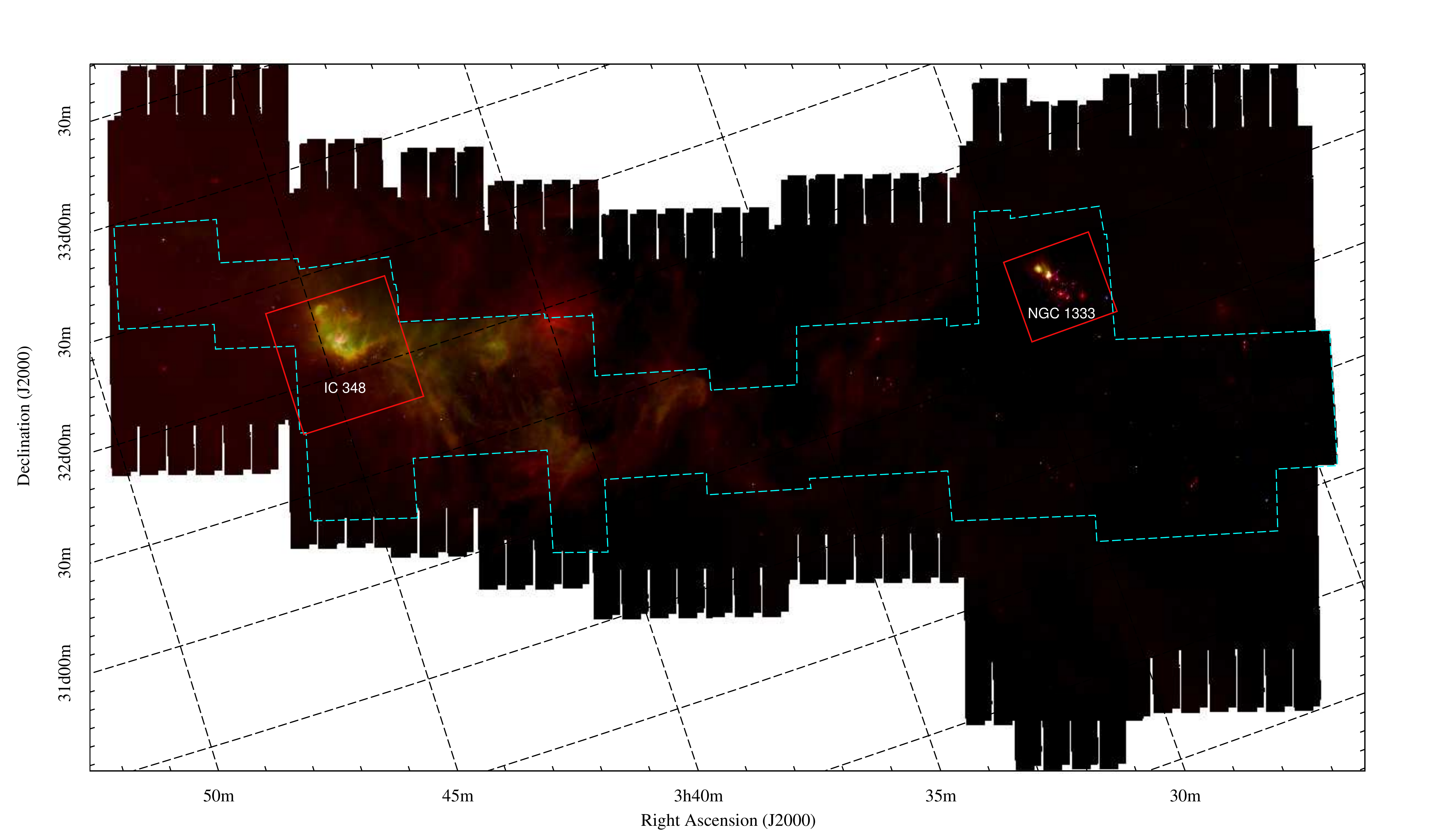}
\caption{Three-color (Blue = 4.5 \micron, Green = 8 \micron, and Red = 24 \micron) mosaic of Perseus. Edges of the image are the MIPS scan region. The IRAC observation area is given by the cyan dashed line. The red boxes indicate the IC 348 and NGC 1333 cluster regions as described in Section 3.}
\label{fig:color}
\end{figure}

The Perseus molecular cloud was mapped with two of the \textit{Spitzer Space Telescope's} instruments, IRAC and MIPS. Figure \ref{fig:color} is a three-color composite map of Perseus from the IRAC 4.5 and 8 $\mu$m and MIPS 24 $\mu$m data. The {\it Spitzer} maps cover 3.86 square degrees in the IRAC observation wavelengths of 3.6, 4.5, 5.8, and 8.0 $\mu$m and approximately 10.5 square degrees with MIPS at 24, 70, and 160 $\mu$m. However, due to the nature of the detector, the 160 \micron\ maps are not complete and provide only limited photometry for a portion of the YSOs. Therefore, our discussion will be limited to data from 3.6 to 70 $\mu$m. Rebull et al. (2007) discussed the 160 \micron\ mosaic. In Figure \ref{fig:color}, the edges of the image are the MIPS scan boundaries and the IRAC observation area is denoted by the dashed lines. We will focus on the area where the two maps overlap, namely the nearly 4 square degrees observed by IRAC.

The outline of the Perseus mapped region was selected using the \av\ = 2 mag contour from \textsuperscript{13}CO extinction maps of \citet{pad99}. The c2d team chose this criteria in order to observe the regions of densest gas and dust in the cloud complex and, by association, observe all of the young and forming stars in Perseus. A correlation also exists between \textsuperscript{13}CO and near-infrared dust extinction \citep{kai11}. \citet{pin08} showed that the \textsuperscript{13}CO column density is correlated with visual extinction (\av) in Perseus for \av\ $<$ 10 mag. However, \citet{hei10} concluded that, while \textsuperscript{13}CO may be a good tracer of gas surface density, it is not linearly correlated with visual extinction in Perseus when considering \av\ up to 25 mag and, using standard conversion factors, may underestimate mass. 

The IRAC maps fit tightly to the \av\ = 2 mag outline, whereas MIPS was able to scan over a much larger area in less observing time due to the design of the instrument. NGC 1333 and IC 348 were mapped separately from the c2d map in conjunction with \textit{Spitzer} Guaranteed Time Observers program \citep[GTOs;][]{gut08,lad06}. The GTO maps have been combined with the larger c2d maps and are included in this analysis. 

The IRAC and MIPS observations of Perseus were discussed in detail in \citet{jor06} and \citet{reb07} respectively. We only discuss pertinent details of the observations here. Each instrument mapped Perseus twice in separate time epochs anywhere from several hours to 7 months apart, depending on the region, to identify and eliminate asteroids and artifacts. The raw data were processed through the c2d data reduction pipeline to create mosaic maps, reduce instrumental effects, and produce a band-merged source catalog for Perseus that includes both Two Micron All Sky Survey (2MASS) and {\it Spitzer} photometry. The fluxes given in the c2d catalog at each IRAC and MIPS band are the average of the single-epoch fluxes \citep{eva07,eva09}. The c2d catalogs for Perseus as well as the other molecular clouds and cores included in the survey are available online through the NASA/IPAC Infrared Science Archive \citep{c2d}. A complete description of the data processing and source extraction for all of the c2d data can be found in the first papers in this series \citep[e.g.,][]{you05,har06} and the final c2d data delivery document \citep{eva07}.

\begin{figure}
\figurenum{2}
\plotone{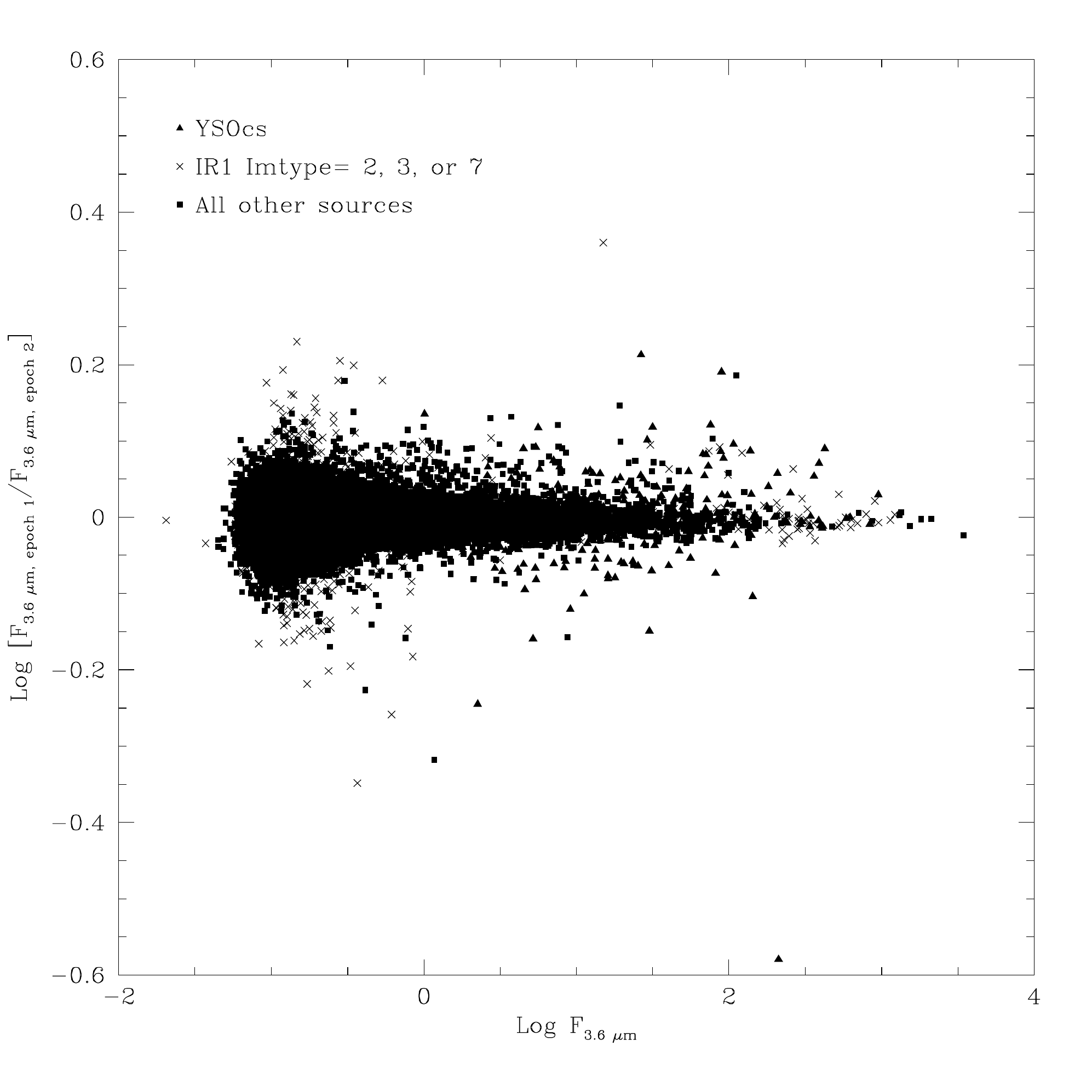}
\caption{The log of the ratio of fluxes in the IRAC 1 band (3.6 \micron) between the two observation epochs versus the log of the 3.6 \micron\ flux in mJy from the combined epochs for all sources where the uncertainty is less than 10\% of the flux in single epoch. Sources identified as YSO candidates by the c2d team are triangles. Crosses represent sources flagged as extended, double, or with a signal-to-noise ratio that was too small to determine shape (Imtype = 2, 3 or 7) in the IRAC 1 map as discussed in Section 2. All other source types, including stars, are squares.}
\label{fig:fluxratio}
\end{figure}

Figure \ref{fig:fluxratio} plots the ratio of observed fluxes at 3.6 $\mu$m (IRAC Band 1) between the two observation epochs versus the 3.6 $\mu$m flux derived from the combined epochs for all sources in the cloud with a single epoch flux to uncertainty ratio greater than 10. The flux measurements are very consistent, differing by less than 15\%, between the epochs, with an average flux ratio of $Log$ $[F_{(3.6 \micron, epoch 1)}/F_{(3.6 \micron, epoch 2)}] = -0.0032$. Figure \ref{fig:fluxratio} shows a slightly larger spread in the flux ratios ($-0.2 <$ $Log$ $[F_{(3.6 \micron, epoch 1)}/F_{(3.6 \micron, epoch 2)}] < 0.2$) for the faintest sources ($F_{3.6 \micron} < 1$ mJy). The few sources with more extreme flux ratios between epochs were mostly flagged in the data processing as extended, double, or with a signal-to-noise ratio that was too small to determine shape \citep[Imtype = 2, 3, or 7;][]{eva07}. These Imtypes are indicated as X's in Figure \ref{fig:fluxratio}, where one can see that the outliers fall mostly in these categories. 

Three objects classified by c2d as young stellar objects (YSOs; see Section 3) have a large ratio of 3.6 \micron\ fluxes between the first and second observations (triangle outliers in Figure\ref{fig:fluxratio}). The variation in flux between the two observation epochs for these objects is likely not physical. One YSO in the IC 348 region was flagged with an uncertainty ratio too low to accurately determine shape (index number 348 in Table \ref{tab:348}). Another YSO in IC 348 (424 in Table \ref{tab:348}) has a relatively low flux at 3.6 \micron, $Log$ $F_{3.6 \micron} = 0.068$, placing it at the edge of the spread in Figure \ref{fig:fluxratio}. This object is also situated in extended nebulosity near a source with a large 3.6 \micron\ flux. The third YSO outlier (209 in Table \ref{tab:1333}) has a combined epoch flux of 39 mJy at 3.6 \micron\ and a $Log$ $[F_{(3.6 \micron, epoch 1)}/F_{(3.6 \micron, epoch 2)}] = 0.213$ for observations 7 months apart. YSO 209 is a Class 0+I object (see Section 4) located in the NGC 1333 region, but again is potentially confused with bright nebulosity. 

\section{YSO Selection}

A major goal of the c2d program was to identify young stellar objects (YSOs) in the large molecular clouds. We follow the method of previous papers in this series for selecting YSO candidates in Perseus, in which YSOs are distinguished from reddened background galaxies using color-color and color-magnitude diagrams in the 2MASS (H and K), IRAC (4.5 and 8 \micron), and MIPS (24 \micron) bands.  This method is described here briefly, however a detailed description can be found in the c2d data delivery document \citep{eva07} and by \citet{har07}.

The c2d data were compared with a galaxy sample in the ELAIS N1 field from the \textit{Spitzer} Wide-area Infrared Extragalactic (SWIRE) catalog \citep{lon03}. ELAIS N1 is well above the galactic plane and not expected to contain any YSOs or molecular clouds. The SWIRE data was used to eliminate confusion of reddened background galaxies with YSOs, however, it does not control for reddened background stars in our own galaxy. IRAC and MIPS maps of the field observed by the SWIRE team were resampled to match the c2d sensitivity and reprocessed in the c2d pipeline so accurate color comparisons could be made. In addition, the SWIRE data were also reddened using the extinction profile of Perseus generated from {\it Spitzer} data (Figure \ref{fig:avhist}). Therefore, galaxies in the SWIRE field would appear reddened as if behind a Perseus-like molecular cloud. The reprocessed SWIRE data were plotted in infrared color-color and color-magnitude spaces to determine the positions of galaxies in the diagrams. This process resulted in cutoff limits in each color-color and color-magnitude space to divide YSOs from galaxies. 

The color-color and color-magnitude diagrams used for YSO selection in Perseus are in Figure \ref{fig:colormag}. The lines in each color-color or color-magnitude space are the cutoffs that define the YSO candidate criterion in the various color-magnitude spaces created by comparison with the SWIRE data and are the same as those presented and discussed by Harvey et al. \citeyearpar{har07}.  The c2d team constructed the cutoffs in the color-magnitude diagrams to be smooth, exponentially decaying probabilities around the lines in Figure \ref{fig:colormag} so that sources far below the cutoffs are assigned the highest probability of being extragalactic. The probability of extragalactic contamination is highest in the center of the ellipses in the $[24]$ versus $[8.0] - [24]$ diagrams and decreases radially.

Perseus sources with infrared colors that were consistent with dust emission rather than extincted background stars were assigned an unnormalized total ``probability'' that the source is extragalactic. This proxy for probability is formed from the product of the individual probabilities from the color-magnitude diagrams in Figure \ref{fig:colormag} and factors based on source properties such as $K-[4.5]$ color, MIPS flux density, and whether it was found to be extended at 3.6 and 4.5 \micron\ \citep{har07}. The cutoffs for $K-[4.5]$ color are also shown in Figure \ref{fig:colormag}. Bluer colors reflect a higher probability that the source is extragalactic.

In Figures \ref{fig:avhist}, \ref{fig:colormag}, and \ref{fig:galhist} and the following text, we divide Perseus into three distinct regions for comparison of the YSO properties. The sources associated with the IC 348 and NGC 1333 clusters are shown separately from the ``Remaining Cloud" (hereafter RC), which includes all the other sources, such as B1 and L1455, associated with the Perseus cloud complex. Figure \ref{fig:color} displays the boundaries of these two cluster regions in red. The NGC 1333 region was defined from 03\textsuperscript{h}28\textsuperscript{m}00\textsuperscript{s} to 03\textsuperscript{h}30\textsuperscript{m}00\textsuperscript{s} in right ascension and +31\degr 06\arcmin 00\arcsec\ to +31\degr 30\arcmin 00\arcsec\ in declination. The IC 348 region runs from 03\textsuperscript{h}43\textsuperscript{m}12\textsuperscript{s} to 03\textsuperscript{h}46\textsuperscript{m}00\textsuperscript{s} in right ascension and +31\degr 48\arcmin 00\arcsec\ to +32\degr 24\arcmin 00\arcsec\ in declination. The RC includes all of the IRAC observation area (dashed outline in Fig. 1) outside of the cluster boundaries. 

These three regions were previously defined in \citet{jor06} and \citet{reb07} based on the boundaries of the GTO observations of the clusters {\citep{gut08,lad06}}. The regions do not represent physical boundaries of the clusters, but reflect {\it Spitzer's} mapping footprint. Therefore, the IC 348 and NGC 1333 regions may include some YSOs that are not physically associated with the clusters. However, for consistency within this series of papers, we maintain these somewhat arbitrary boundaries. 

Figure \ref{fig:galhist} is a histogram of the number of identified sources with infrared flux versus the ``probability'' of being a background galaxy. Sources with low probabilities were considered YSO candidates. The cutoff to separate YSOs and galaxies in terms of probability is $Log$ P $\leq$ $-$1.47 (as in \citet{har07}).  There is a small tail of sources found in Perseus between $Log$ P = $-$1.5 and $-$2.0. Harvey et al. \citeyearpar{har07} predicted no more than zero to one contaminant galaxies in this tail based on comparisons with the SWIRE catalog. Most of the extragalactic sources appear in the RC histogram rather than being associated with the young clusters in Perseus because of their uniform spacing over the cloud area. 

Previous papers in this series have reported the number of YSO candidates (YSOc) in a given c2d cloud. However, in recent years, the c2d YSOc lists have been used and refined in a number of studies (e.g., E09, \citealt{dun13}). In the course of these studies, all of the YSOc were visually inspected to eliminate resolved galaxies and artifacts (E09). Therefore, following the terminology in E09, the ``candidate'' designation has been dropped for this study and we will discuss the YSOs found in Perseus.

Using the described selection method, we identified 385 YSOs in Perseus. IRAC and MIPS fluxes for these sources, as well as the YSOs in the other c2d clouds, were published in online tables as part of a paper summarizing the c2d \textit{Spitzer} Legacy project results (E09). The tables also provide additional SED data, classifications, and derived properties, such as bolometric temperature and luminosity, for all of the Perseus YSOs. Dunham et al. \citeyearpar{dun13} updated the SEDs and other properties of the protostellar population in Perseus.

Recently, \citet{hl13} developed a new method for separating YSOs and galaxies in \textit{Spitzer} data sets using multi-dimensional (multi-D) magnitude space. They analyzed all of the c2d clouds with this new method, adding 100 YSO candidates and removing 16 YSOs in Perseus compared to the E09 c2d list for a total of 469 YSOc in Perseus. Figure \ref{fig:hlmap} shows the location of the \citet{hl13} new YSO candidates on the 8 \micron\ map of the cloud. Sixteen c2d-identified YSOs were not included by \citet{hl13} for various reasons (see also their Table 5): 6 are jet knots, 6 appear to be confused with nearby sources or cloud structures, 3 were not identified as high reliability sources by c2d, and one was identified by the multi-D process as a giant star. We agree with Hsieh and Lai's characterization of these 16 sources and have removed them from our list of YSOs in Perseus.

The multi-D method developed for identifying YSO candidates by \citet{hl13} is relatively new and completely different from the established c2d method presented in previous papers in this series and in other works by the c2d team. Because our method, as described above, does not identify the 100 additional sources in the \citet{hl13} candidate list as YSOs and for consistency with the previous work from the c2d team, we do not include the new \citet{hl13} YSO candidates. The \citet{hl13} candidates likely include previously unidentified YSOs in Perseus. However, we suggest further observations, such as optical spectra or maps of the molecular cores, are needed to confirm the classification of these faint sources and rule out possibility that the candidates are reddened AGB stars or background galaxies. Therefore, we present 369 YSOs identified by the c2d method, 385 from E09 minus 16 excluded by \citet{hl13}, in Perseus. 

E09 calculated the star formation rate (SFR) for all of the c2d clouds based on the number of YSOs (N$_{YSO}$) identified in each cloud. Assuming a mean mass per YSO of 0.5 M$_{\sun}$ and a period of star formation of 2 Myr, E09 report a SFR in Perseus of 96 M$_{\sun}$ Myr$^{-1}$. With 369 instead of 385 YSOs and a slightly lower SFR of 92 M$_{\sun}$ Myr$^{-1}$, Perseus maintains the highest SFR among the c2d team's molecular clouds. If all of YSO candidates from \citet{hl13} are included, the SFR in Perseus increases to 117 M$_{\sun}$ Myr$^{-1}$.

Figure \ref{fig:hlmap} shows the distribution of the 369 YSOs identified in Perseus. YSOs are found in groups and clusters associated with known regions of star formation. We list and identify the YSOs in each cluster region and the RC. Tables \ref{tab:348} and \ref{tab:1333} list the 143 YSOs in the IC 348 region and 104 YSOs in the NGC 1333 region respectively. Table \ref{tab:rc} identifies 122 YSOs in the RC. 

\section{YSO Classification}

All of the YSOs identified in the c2d molecular cloud survey with sufficient data were placed in one of four classes according to their spectral index, $\alpha$: Class 0+I, Flat (i.e., flat-spectrum), Class II, and Class III. We have reclassified the YSOs in Perseus according to an updated spectral index. Therefore, some sources are classified differently in this work than in E09.

The spectral index is defined by a least squares fit to the photometry of the source between 2 and 24 $\mu$m according to Equation 1 in E09. The near-infrared flux data are from 2MASS. We follow the class definitions of \citet{gre94}, where $\alpha \geq$ 0.3 is a Class I,$- 0.3 \leq \alpha < 0.3$ is a Flat source, $-$ 1.6 $\leq \alpha < -0.3$ is a Class II, and $\alpha < -1.6$ is a Class III.

Because \citet{gre94} does not distinguish between Class 0 and Class I sources and the classes cannot be separated based on spectra index (\citealt{dun14}; M. M. Dunham et al. 2015, in preparation), we classify all sources with $\alpha \geq$ 0.3 as Class 0+I.  In addition, the \citet{gre94} classification scheme does not put a lower limit on $\alpha$ for Class III sources and therefore includes YSOs that with spectral energy distributions that resemble bare stellar photospheres.  As noted by \citet{gut09}, {\it Spitzer} data alone cannot identify diskless Class III sources with no infrared excess. Therefore, our Class III census represents a lower limit for Perseus.

Evans et al. (2009) corrected the c2d large cloud photometry for extinction. Using a mean extinction of A$_{\textnormal{v}}$ = 5.92 in Perseus, E09 classified their 385 YSO candidates: 76 Class 0+I, 35 Flat, 244 Class II, and 30 Class III. Following E09, we assign extinction values to all YSOs as follows:
\begin{enumerate}
\item We adopt extinction values from the literature for Class II and III YSOs (classified via infrared spectral index; see E09 for details) included in published optical studies.
\item We de-redden the remaining Class II and III YSOs to the intrinsic near-infrared colors of an assumed spectral type of K7, found to be fairly representative of the majority of Class II and III YSOs in the c2d clouds (\citealt{oli09,oli10}; see also E09 for details).
\item We de-redden all of the Class 0+I and Flat spectrum YSOs (again classified via infrared spectral index) using the mean extinction toward all Class II YSOs in Perseus (\av\ = 5.9 mag). The mean extinction is used since we cannot directly determine the line-of-sight extinction from the cloud for these sources.
\end{enumerate}

Once the extinction values are assigned, we use these values combined with the \citet{wd01} extinction law for $R_V = 5.5$ to correct the photometry for extinction.  The choice of the $R_V = 5.5$ law rather than the $R_V = 3.1$ law is motivated by several studies showing that the former is more appropriate for the dense regions in which stars form (e.g., \citealt{cha09}).  While the original extinction corrections in E09 only used the absorption cross section, here we adopt the sum of the absorption and scattering cross sections in order to correct for the total line-of-sight extinction.

Tables \ref{tab:348} - \ref{tab:rc} list the absorption and scattering corrected spectral indices and photometry for each YSO. The 369 YSOs identified in this work were reclassified according to their newly corrected spectral index. We find 70 Class 0+I, 32 Flat, 231 Class II, and 36 Class III sources in Perseus. Although the number of sources in each class differs from E09, the fractions of each class compared to the total number of YSOs is not significantly different from those reported by E09 in comparing the five c2d clouds. 

Perseus has a larger fraction, 19\%, of Class 0+I sources compared to the other c2d star forming clouds. \citet{hl13} classified their 469 YSOc in Perseus using the same class definitions, reporting 99 Class 0+I, 49 Flat, 272 Class II, and 49 Class III sources (their Table 8). They find more YSOc in each category compared to the c2d list, however the fractions of each class remain very similar to this work. Namely, \citet{hl13} also find a large fraction, 21\%, of Class 0+I sources. Table \ref{tab:class} gives the fraction of the total number of YSOs for each classification.

These observational YSO classifications are suggestive of physical stages of star formation. \citet{hei15} used HCO\textsuperscript{+} observations to classify Class 0+I and Flat sources identified in the c2d and Gould Belt (M. M. Dunham et al. 2015, in preparation) {\it Spitzer} surveys of molecular clouds as physically embedded Stage 0+I \citep{rob06} objects that retain envelope material. They find 72\% of the {\it Spitzer} Class I and 48\% of Flat sources are also Stage 0+I protostars.

YSO classification has also been connected to the physical nature of disks around young stars. \citet{lad06} and \citet{tei12} used $\alpha$ derived from IRAC data to place YSOs with optically thick disks and anemic disks into distinct categories. The thick and thin disk classifications developed by \citet{lad06} are very similar to the Class II and Class III definitions of \citet{gre94}. Using these definitions, we find 228 YSOs with thick disks (-0.5 $> \alpha_{IRAC} \geq$  -1.8) and 29 with anemic disks (-1.8 $> \alpha_{IRAC} \geq$ -2.56) in Perseus. The IC 348 cluster region contains nearly half of these sources (126) and the majority of the thick disks (59\%) in the cloud.  

\section{YSO Distribution}
\subsection{Across the Cloud}

More YSOs were identified in Perseus than in any of the other clouds studied in the c2d survey. However, it also covers the most area. Perseus, as previously reported in E09, has a density of 5.0 YSOs per square parsec, placing it behind Serpens and Ophiuchus in terms of number of YSOs per area. Table \ref{tab:class} lists the number of YSOs per square parsec, assuming a distance of 250 pc, for each region of Perseus. If we adopt the distance found by \citet{hir08} of 235 pc for NGC 1333, the density increases to 5.7 YSOs pc$^{-2}$.

As noted in Section 3, the YSOs in Perseus are not evenly distributed across the cloud (Figure \ref{fig:hlmap}). Instead there are two main sections with star formation activity. To the east, lie B5, IC 348, and L1468. NGC 1333, B1, L1448, and L1455 are on the opposite end of the mapped region (Figure \ref{fig:regions}). The two clusters, IC 348 and NGC 1333, contain 67\% of the YSOs in Perseus. NGC 1333 has the highest density of YSOs, 32 pc$^{-2}$. Many of the 122 YSOs in the RC are associated with the other known star-forming associations such as B1 (22), L1448 (5), and L1455 (11). L1455 has the lowest density of all the regions at 1.4 YSOs pc$^{-2}$.

Only two YSOs are found in the middle of Perseus far from the clusters and known star-forming associations (see Figure \ref{fig:regions}): index numbers 304 and 305 in Table \ref{tab:rc} and E09. YSO 304 is within a few arcmin of the dark cloud L1468. However, there is no evidence in the literature of star-formation activity in L1468. Both YSOs are found toward regions where the visual extinction is greater than 4 magnitudes and both are Class II sources. \citet{hl13} identified three additional YSO candidates (their numbers 227, 228, \& 231) in the mid-cloud region. One of their candidates (231) is within 2 arcmin of our YSO number 305. Section 6 discusses the mid-cloud region in terms of visual extinction and star formation.

\subsection{By Class}

\citet{hl13} divided Perseus into eastern and western regions and concluded that western Perseus (including NGC 1333) has the greatest fraction of embedded sources (Class 0+I) of any of the c2d molecular clouds.  The {\it Herschel Space Observatory} observed more infrared emission in its shorter wavelength bands (70 and 160 \micron) in eastern Perseus (IC 348) than the western end of the cloud indicating that the eastern end is warmer \citep{dif13}. Further analysis of the {\it Herschel} data showed regions in the western end of the cloud (NGC 1333 and L1448) have the greatest fractions of Class 0+I YSOs \citep{sad13}. 

Figure \ref{fig:regions} shows the locations of YSOs according to their classifications overlaid on the 8 \micron\ IRAC map. Named star-forming associations are labeled and the clusters and larger associations (B1 and L1455) used for analysis of the YSO populations in this work are outlined in the figure. Table \ref{tab:class} lists the regional number and fraction of YSOs for each class in the clusters, RC, B1, L1455, and the eastern and western sections of the clouds as well as the ratio of Class II to Class 0+I YSOs for the region. The charts in Figure \ref{fig:pie} are a visual representation of the distribution of YSOs by class for regions listed in Table \ref{tab:class}.

Our results agree with those of \citet{hl13} and the {\it Herschel} studies \citep{dif13, sad13}. By dividing the Perseus cloud into eastern and western sections along a line of right ascension at 3\textsuperscript{h}38\textsuperscript{m}00\textsuperscript{s} (Fig. \ref{fig:regions}), we find three-quarters of the Class 0+I and Flat YSOs in the western section. The eastern section, an area of approximately 1.6 square degrees, includes IC 348 and contains the majority of the more evolved sources, 61\% of Class II and 81\% of Class III. 

Assuming a constant birthrate of stars, the ratio of the number of Class II to Class 0+I YSOs is used as a relative age estimate between star-forming regions \citep{gut09}. Class II sources are more abundant than Class 0+I YSOs in Eastern Perseus by a ratio of 9.3 to 1. When we analyze the YSO population in the clusters, the Class II/(0+I) ratio is 9.5 in IC 348 but only 2.1 in NGC 1333, implying IC 348 is older. \citet{gut09} found ratios of 9.0 and 2.7 for IC 348 and NGC 1333 respectively and both the separate IRAC \citep{jor06} and MIPS \citep{reb07} studies of Perseus agree that IC 348 contains a larger ratio of Class II to Class I sources. The Class II/(0+I) ratios are consistent with previous age estimates for the clusters (NGC 1333 $<$ 1 Myr \citet{wil04} and IC 348 approx. 2 Myr \citet{lad06}). 

We find more than twice the number of Class 0+I and Flat sources are associated with NCG 1333 as are associated with IC 348, as first suggested by the analysis of IRAC data alone \citep{jor06}. J\o rgensen et al. \citeyearpar{jor06} found that the majority of these more embedded sources were in the smaller star-forming regions in the RC, suggesting that these associations are the sites of the most recent star formation activity in the cloud. However, when Rebull et al. (2007) analyzed the MIPS data, they showed that the abundance of bright Class I and Flat sources is not higher in the RC, suggesting that the J\o rgensen et al. \citeyearpar{jor06} result applies only to faint YSOs.

With the combined IRAC and MIPS data, as well as the extinction corrected source classifications, we find 40 Class 0+I and Flat sources in the defined NGC 1333 region and 46 in the RC. Further, we find the RC has a greater number of Class 0+I and Flat YSOs (28) with MIPS 24 \micron\ fluxes greater than 100 mJy compared to NGC 1333 (15). The Class II/(0+I) ratio in NGC 1333 (2.1) is nearly equal to that of the RC (2.2). Based solely on source counts in these regions, there is not an obvious implied age difference between NGC 1333 and the RC. 

However, individual star-forming associations in the RC do contain higher fractions of Class 0+I sources than NGC 1333, suggesting a younger age for these regions and in agreement with \citet{jor06}. For example, L1448, at the far western edge of the {\it Spitzer} map, contains only embedded YSOs, 5 Class 0+I sources. 

We defined large regions (0.4 square degrees, Fig. \ref{fig:regions}) for the largest star-forming associations outside of the clusters, B1 and L1455. The B1 area is the same as the SCUBA 850 \micron\ map of \citet{sad13}. Nearly half (45\%) of the YSOs in the B1 and L1455 areas are Class 0+I sources in contrast with 27\% in NGC 1333. The ratio of Class II to Class 0+I sources is only 0.8 in B1 and L1455. However, we caution that even with the broadly defined B1 and L1455 regions the total number of YSOs in these regions is relatively small (33). 

We find Class III sources are most common in IC 348 in agreement with \citet{jor06}. 64\% are found in the cluster. The RC contains only a few more Class III sources (8) than NGC 1333 (5). As noted previously, this is a lower limit for Class III sources in these regions since our data cannot identify young objects with star-like spectral energy distributions. Extensive X-ray observations or spectroscopic surveys of the Perseus cloud are needed to find the Class III sources missed by this survey.

\citet{reb15} cross-referenced a large number of NGC 1333 catalogs, including the {\it Spitzer} catalog, and classified YSOs with the combined data sets. Rebull identified more than twice as many YSOs (277) associated with the cluster than this work (104). The Rebull catalog encompasses a larger region than we define for NGC 1333, so some YSOs in \citet{reb15} are in our RC region. The YSOs in Rebull's catalog that we do not identify in this work were generally not detected in most or all of the {\it Spitzer} bands and were identified as YSOs with submillimeter, spectroscopic, or X-ray data. \citet{reb15} lists a large number of Class III sources in NGC 1333 that were identified as young by the Young Stellar Object Variability (YSOVAR) project \citep{reb14} with X-ray detections. 

All of the NGC 1333 YSOs presented here are contained in the \citet{reb15} catalog and the classifications are generally consistent. \citet{reb15} classifies YSOs using the spectral index definitions from \citet{gre94} as in Section 4, however, $\alpha$ is defined from the observed SED without any reddening correction. Rebull places 15 YSOs in a different class than this work due to slightly greater values of $\alpha$. For example, 7 of our Class II sources are placed in the Flat category by \citet{reb15}. The \citet{reb15} fractions of Class I and Flat sources in the cluster are consistent with this work. However, Rebull finds a much larger fraction of Class III sources in NGC 1333, 29\%, because of the greater number of Class III sources identified with X-ray data.

\section{Extended Emission}

Several studies have placed the large scale nebulous emission in the IRAC 8 \micron\ and MIPS 24 \micron\ maps (Figs. \ref{fig:color}, \ref{fig:hlmap}, and \ref{fig:avmap}) of Perseus in the context of star formation and related phenomena. The c2d team compared the IRAC and MIPS extended emission with areas of high extinction \citep{jor06,reb07}. They found no clear correlation at shorter wavelengths; however, the 160 \micron\ mosaic revealed emission that matched extinction maps of the cloud \citep{reb07}. More recently, the COMPLETE (Coordinated Molecular Probe Line Extinction and Thermal Emission) team used the {\it Spitzer} maps of Perseus to help identify shells and bubbles in the cloud as part of their CO survey \citep{arc11}. They found that some of the arc-like nebulosities seen in the IRAC and MIPS maps are related to the edges of shells created by stellar winds.

Figure \ref{fig:avmap} shows \av\ contours overlaid on the MIPS 24 \micron\ map of Perseus. The contours are derived from an extinction map produced by the c2d team using 2MASS and {\it Spitzer} data. The visual extinctions toward sources classified as stars by c2d (see Figure \ref{fig:avhist}) were convolved with a Gaussian beam to create extinction maps of the entire cloud \citep{eva07}. While there are bright regions of 24 \micron\ emission at high extinction, many areas of high extinction (\av\ $>$ 8 mag), such as B1-E, show little to no extended emission.

Furthermore, the IRAC and MIPS extended emission is not necessarily correlated with star formation in Perseus. Most notably, the middle region of Perseus, between 3\textsuperscript{h}37\textsuperscript{m}00\textsuperscript{s} to 3\textsuperscript{h}41\textsuperscript{m}00\textsuperscript{s}, contains only two Class II YSOs (see Section 5) despite bright extended emission at 8 and 24 \micron\ (Figures \ref{fig:hlmap} and \ref{fig:avmap}).

CO and infrared maps show extended emission and some areas of high extinction in this middle region \citep[e.g.,][]{rid06a}. The region also contains dark clouds such as L1468 and B1-E (Figure \ref{fig:regions}). However, there has been little evidence found of star formation, suggesting that the structure in the middle region is perhaps not filamentary enough to form stars \citep{dif13}. The COMPLETE survey highlights the lack of \textsuperscript{13}CO in the mid-region of Perseus as corresponding to a shell, CPS 5, created by a nearby O or B star \citep{goo09, arc11}. The size and location of CPS 5 is indicated on Figure \ref{fig:avmap} against the 24 \micron\ image of Perseus. 

\citet{eno06} mapped Perseus at 1.1 mm and found no dense clumps in the middle of Perseus. Sadavoy et al. (2012) studied the B1-E association, located to the east of NGC 1333 and B1, toward the midsection of the cloud as shown in Figure \ref{fig:color}, as part of the {\it Herschel Space Observatory} Gould Belt survey and found it to contain dense cores but no young stars. Figure \ref{fig:avmap} shows high extinction (mean \av\ $\approx$ 10 mag) toward B1-E, however, we do not find any YSOs in the region. \citet{hl13}, did report one YSO candidate in the region (their 227) using the c2d data.

It has been further postulated that Perseus may consist of two separate clouds that are not physically associated \citep[e.g.,][]{kir06,eno06} or possibly a chain of clouds at a range of distances (250 to 350 pc) along the line of sight \citep{arc11}. If this is the case, the central region of Perseus as seen from Earth may be a physical gap between clouds. These scenarios would naturally explain the lack of YSOs in the mid-regions of Perseus. However, following the lead of earlier papers in this series and other recent surveys of Perseus mentioned here, we continue to assume the entire mapped region to be associated and at the same distance, while acknowledging the complexities of such a large cloud.

\section{Cloud Structure and Relation to Star Formation}

\subsection{Extinction Patterns}

\begin{figure}
\figurenum{3}
\plotone{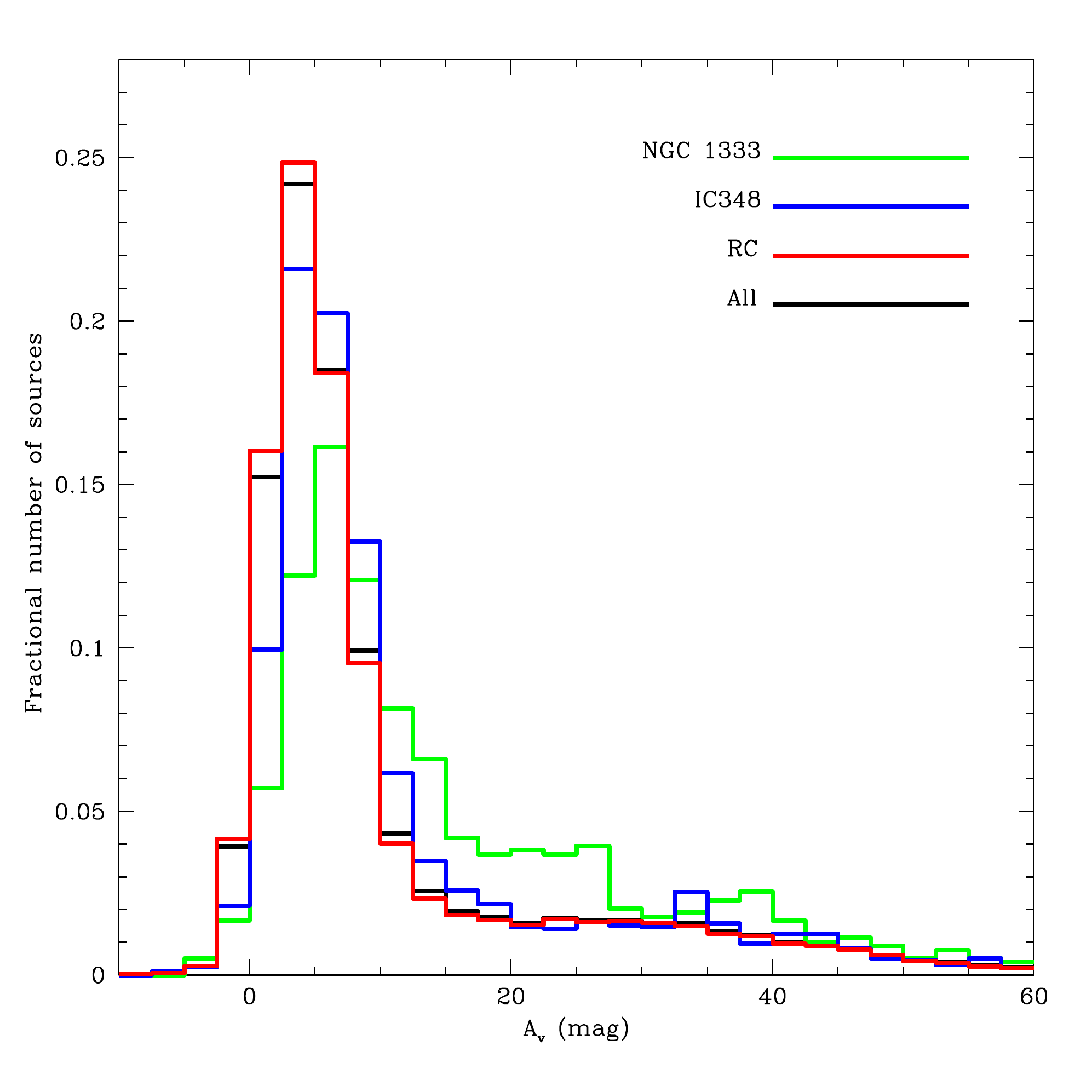}
\caption{The distribution of the visual extinctions found toward the sources classified as stars in Perseus observed with both IRAC and MIPS. The black line is the distribution toward all stars. The distributions for stars toward NGC 1333, IC 348, and the RC are shown as green, blue, and red lines respectively.}
\label{fig:avhist}
\end{figure}

\begin{figure}
\figurenum{4}
\plotone{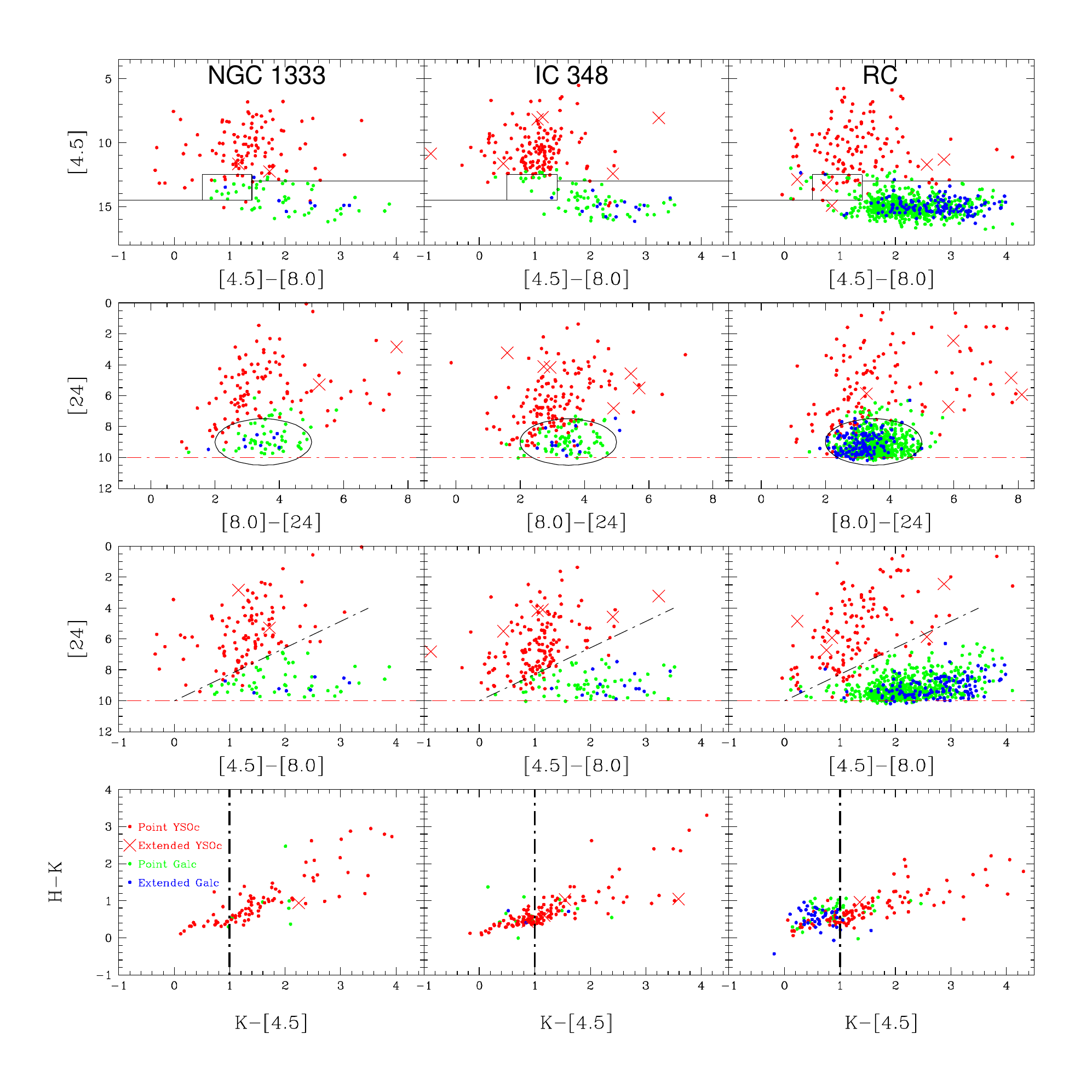}
\caption{Color-color and color-magnitude diagrams for the IC 348, NGC 1333, and the RC regions used to identify extragalactic sources and thus identify YSOs in Perseus. H and K-band data are from 2MASS. Lines and ellipses are cutoffs that differentiate galaxies and YSOs. Sources below the lines and inside the ellipses are assigned a higher probability of being extragalactic. Red points and crosses indicate sources that were identified as point and extended YSOs. Green and blue indicate point-like and extended extragalactic sources respectively.}
\label{fig:colormag}
\end{figure}

\begin{figure}
\figurenum{5}
\plotone{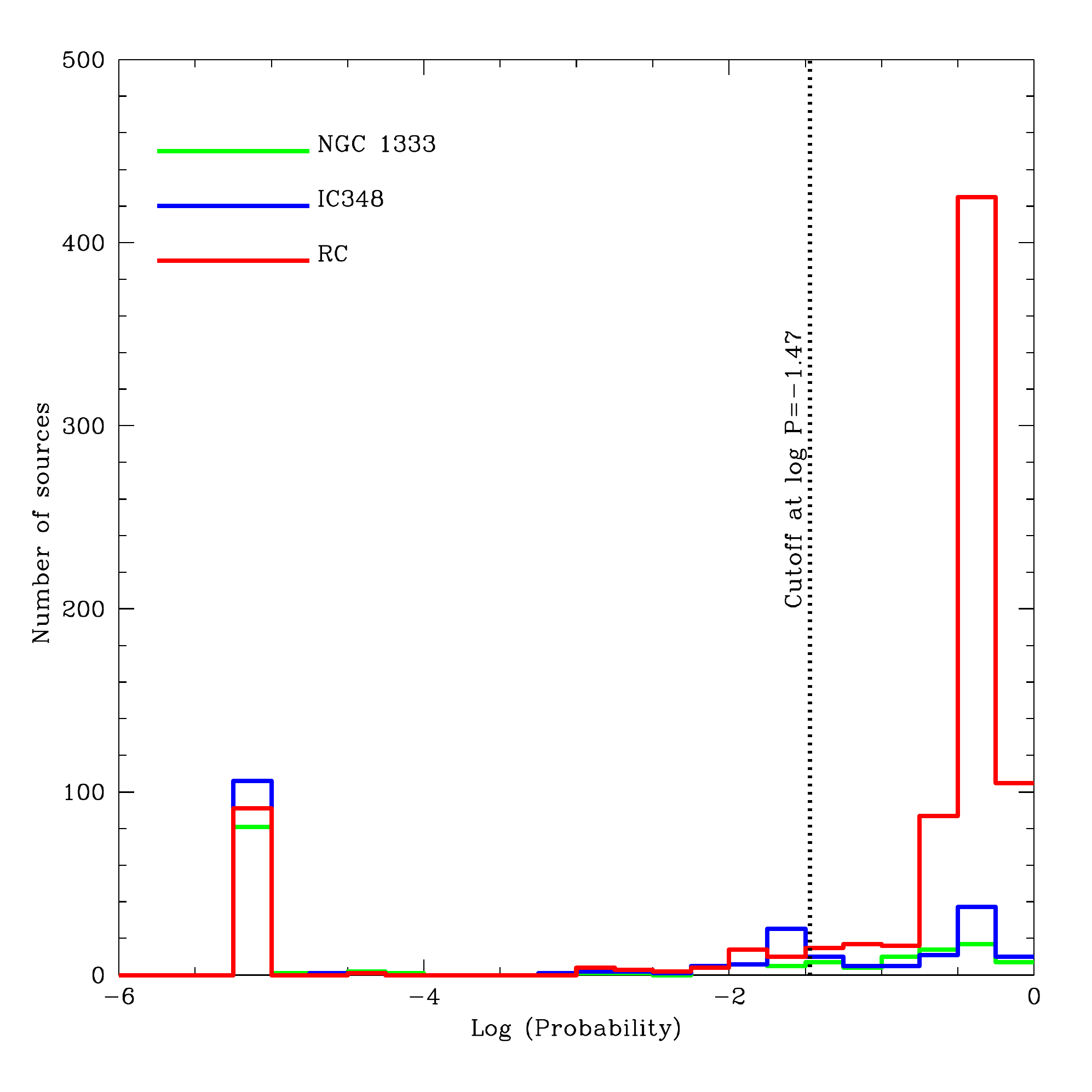}
\caption{Plot of the number of sources versus the unnormalized total ``probability'' of being an extragalactic source for Perseus. This proxy for probability is formed from the product of the individual probabilities from the color-magnitude diagrams in Figure \ref{fig:colormag} and other factors as described in Section 3. The green, blue, and red lines are the probability distributions toward NGC 1333, IC 348, and the RC respectively. The vertical dashed line is the cutoff to separate YSOs and extragalactic sources at P = $-$1.47. }
\label{fig:galhist}
\end{figure}

\begin{figure}
\figurenum{6}
\plotone{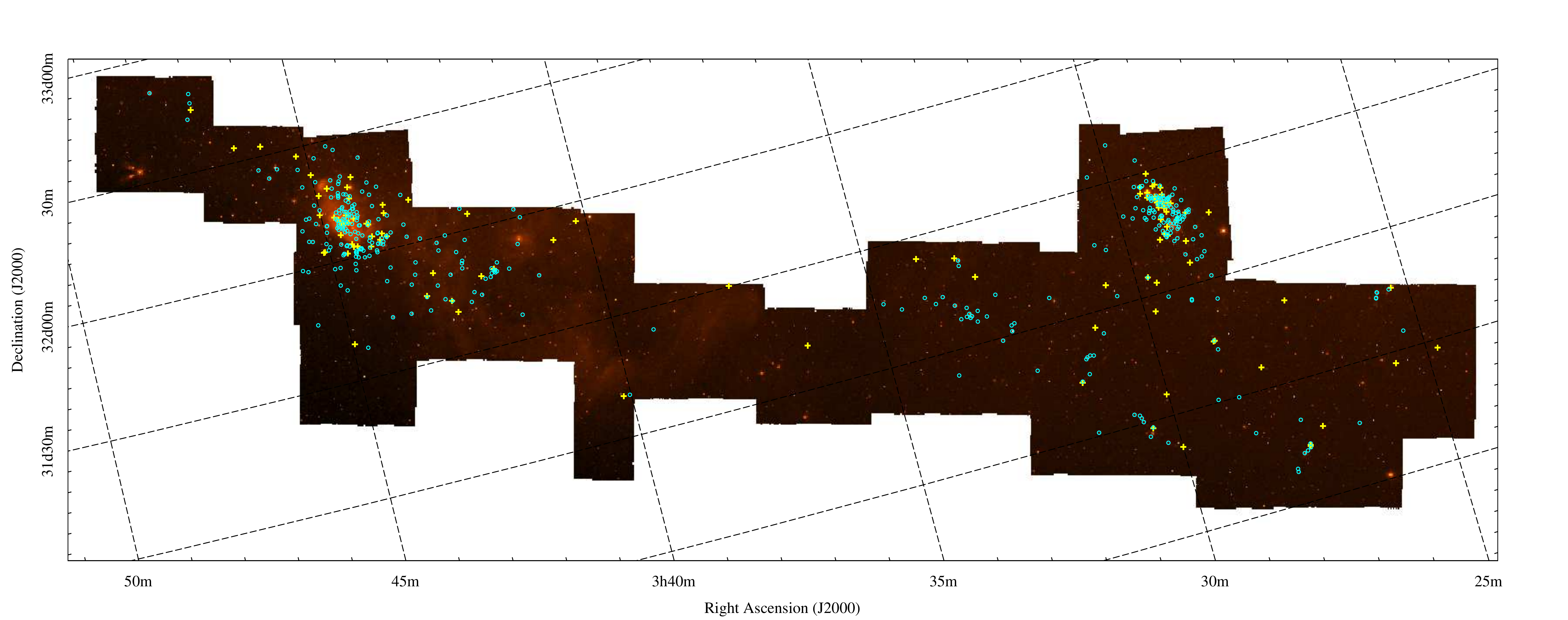}
\caption{The 369 YSOs identified in this work (cyan circles) and the new candidate YSOs from Hsieh \& Lai (2013, yellow crosses) on the IRAC 4.5 \micron\ map of Perseus.}
\label{fig:hlmap}
\end{figure}

\begin{figure}
\figurenum{7}
\plotone{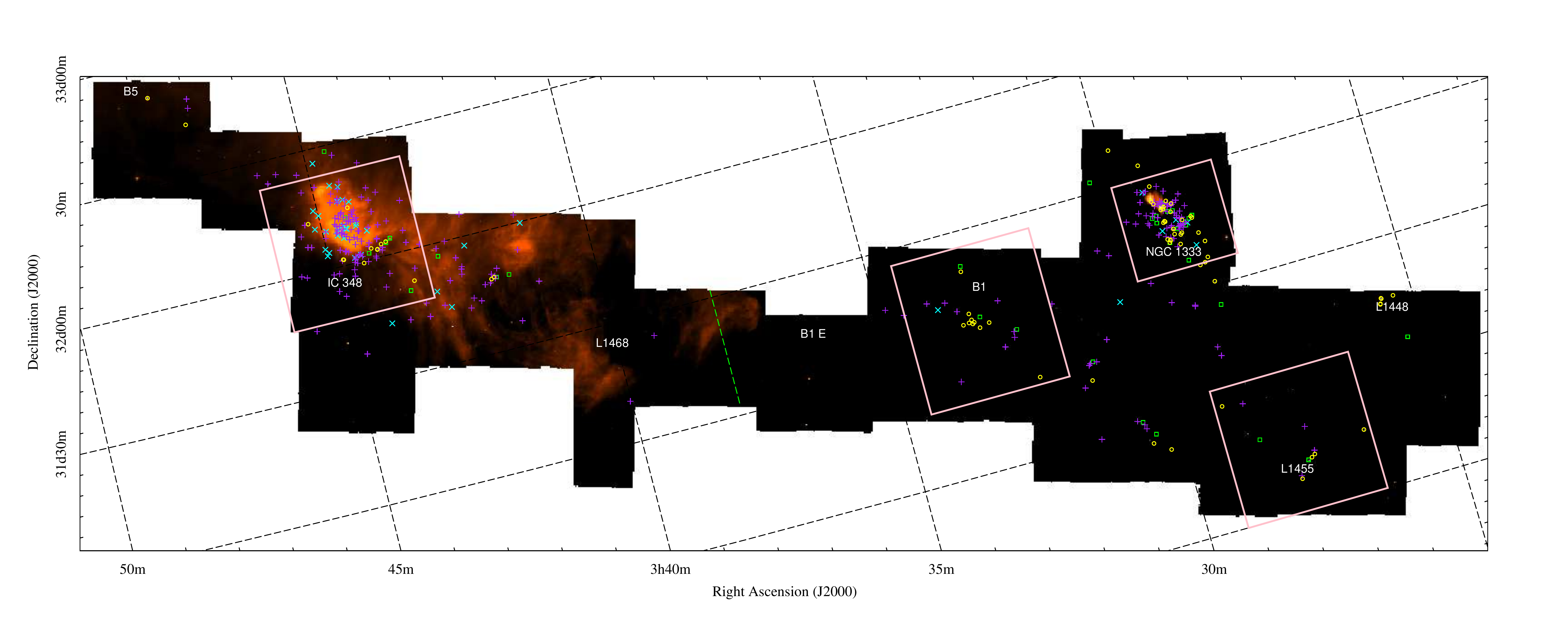}
\caption{IRAC 8 \micron\ map of Perseus. Named regions are labeled. Pink boxes indicate the IC 348, NGC 1333, B1, and L1455 regions described in the text and listed in Table \ref{tab:class}. The green dashed line denotes the right ascension used to divide the cloud into Eastern and Western regions in Section 5. The positions of the classified YSOs are indicated as follows: Class 0+I = yellow circle, Flat = green box, Class II = purple cross, Class III = cyan X.}
\label{fig:regions}
\end{figure}

\begin{figure}
\figurenum{8}
\plotone{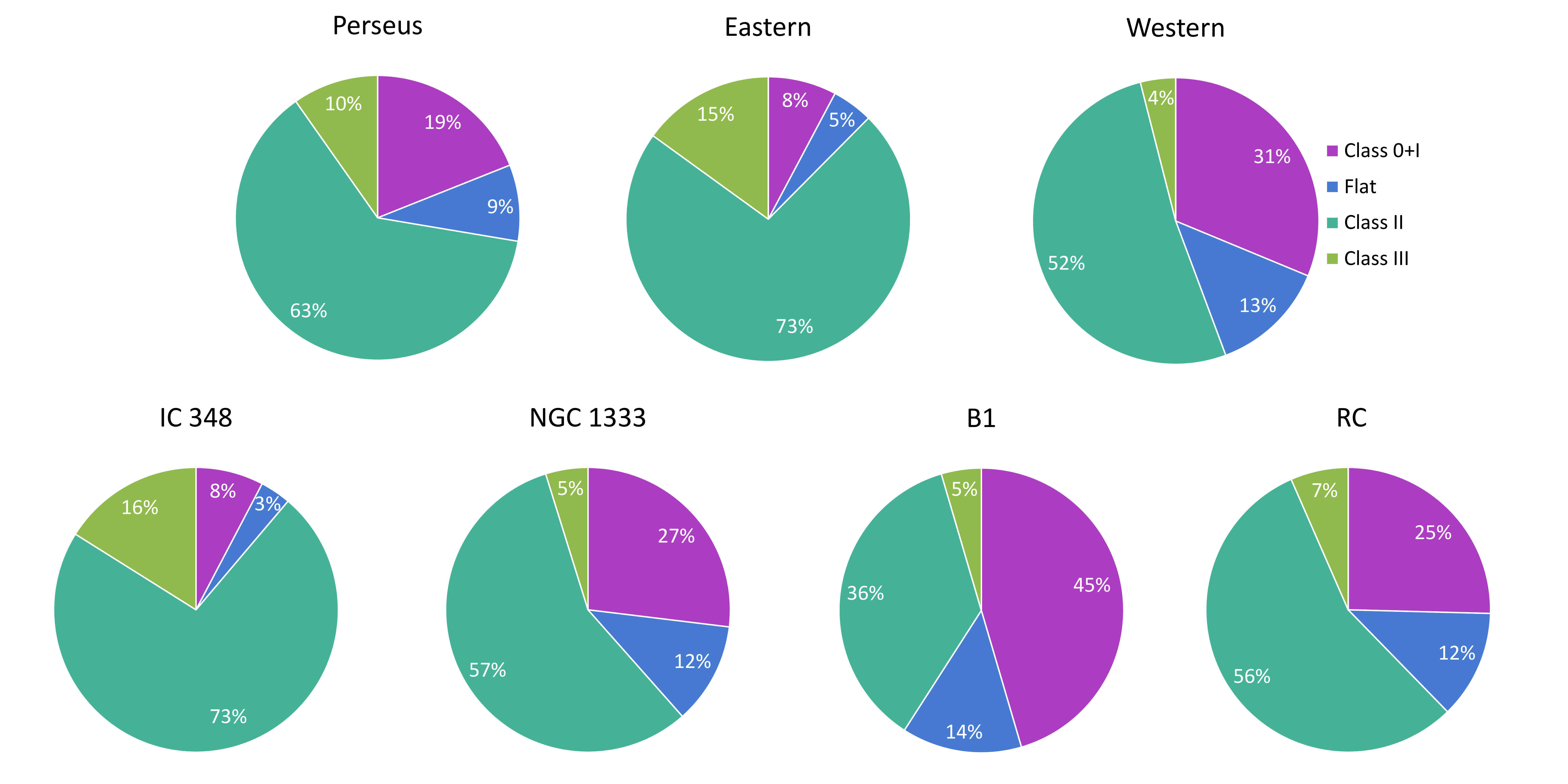}
\caption{Fraction of YSOs in each region by classification. Colors indicate class: Class 0+I = purple, Flat = blue, Class II = teal, and Class III = green.}
\label{fig:pie}
\end{figure}

\begin{figure}
\figurenum{9}
\plotone{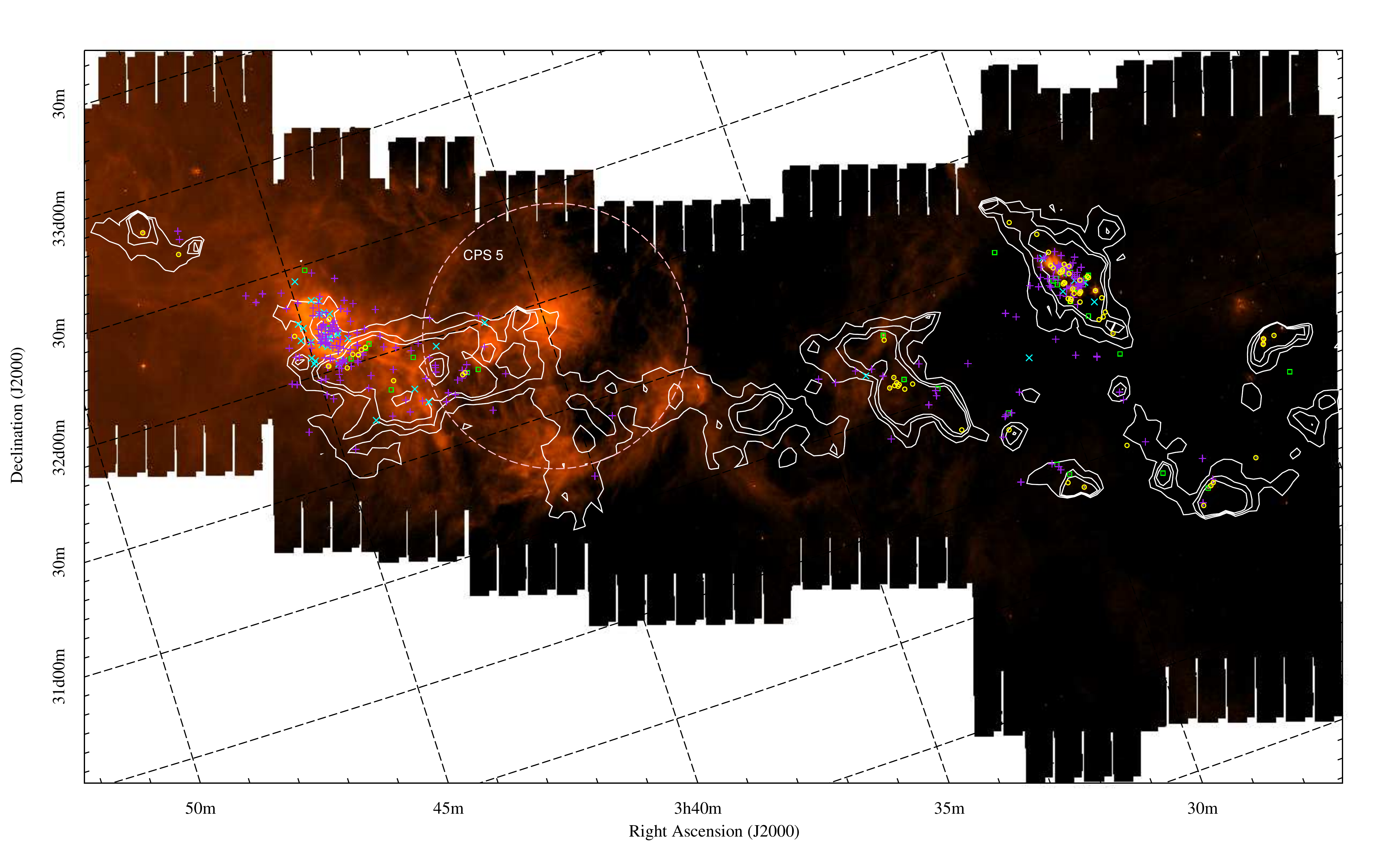}
\caption{ 24 \micron\ map of Perseus overlaid with \av\ = 4, 6, and 8 mag contours derived from 2MASS and {\it Spitzer} data. The positions of the classified YSOs are indicated as follows: Class 0+I = yellow circle, Flat = green box, Class II = purple cross, Class III = cyan X. The size and location of the CO shell CPS 5 discussed in Section 6 is indicated by the large dashed circle.}
\label{fig:avmap}
\end{figure}

Histograms of visual extinction (\av) towards IC 348, NGC 1333, the RC, and all of Perseus using stars observed with IRAC and MIPS are presented in Figure \ref{fig:avhist}. The \av\ was measured along lines of sight though the cloud using the 2MASS and {\it Spitzer} spectral energy distribution of sources classified as stars and adopting the \citet{wd01} extinction law, R$_{\textnormal{v}}$ = 5.5 mag \citep{eva07}. The histograms are presented as a fractional number of stars versus \av. In this work, we limited our data collection by artificially setting the edge of the cloud at the \av\ = 2 mag contour. 

The histograms peak near \av\ = 5 mag and have a tail to higher extinction. Many authors \citep{sch05,rid06a,rid06b,goo09,kai09,fr10,sad14} have presented extinction histograms for Perseus using 2MASS, IRAS, and {\it Herschel} infrared, as well as CO, data in comparison to log-normal probability distribution functions. Log-normal distributions at low column densities suggest the importance of turbulence in the cloud structure (e.g., \citealt{ost01}). Several of the studies (e.g., \citealt{kai09, fr10, sad14}) also show tails above log-normal distributions at higher \av\ ($\geq$ 2 - 5 mag), implying that perhaps self-gravity becomes more important as clouds begin to form stars \citep{kle00,fed08}. Because of the \av\ = 2 mag cutoff in our data, the histograms in Figure \ref{fig:avhist} represent mostly the higher \av\ tail of the distribution. 

We find mean \av\ values of 10.3, 11.7, 15.5, and 10.1 mag for all of Perseus, IC 348, NGC 1333, and the RC respectively. The histogram of extinction for NGC 1333 (green in Fig. \ref{fig:avhist}) peaks at a higher \av\ of 7.5 mag, has a larger mean \av, and a larger tail to high extinction than the other regions. These features indicate different physical conditions in NGC 1333 and IC 348 \citep{goo09} and point to the recent star formation history of NGC 1333.

\subsection{Threshold for Star Formation}

Recent studies of molecular clouds, including Perseus, have suggested an extinction threshold for low-mass star formation (e.g., \citealt{eno08}, \citealt{kir06}, \citealt{hei10}, \citealt{lad10}, \citealt{lad13}, \citealt{eva14}; see also Section 8 of the recent review by \citealt{dun14}). This threshold ranges from \av\ = 4 - 7 mag, below which very few or no dense cores are found. \citet{kir06} found submillimeter clumps in Perseus only where \av\ $\geq$ 5 - 7 mag and suggested a threshold of \av\ = 5 mag linked to magnetic support of the clump, where there is only enough time for ambipolar diffusion if extinction is near the threshold extinction. Figure \ref{fig:avmap} shows our classified YSOs as well as \av\ = 4, 6, and 8 mag contours from the c2d extinction map (see Section 6) on the 24 \micron\ mosaic of Perseus.

Using the background star extinctions averaged over 270\arcsec\ as in \citet{hei15}, we determined the \av\ at the source position for each Class 0+I and Flat YSO, since these extinctions cannot be determined directly. Extinction values for Class II and III sources were previously collected from the literature in \citet{eva09}. 56\% of all the YSOs in Perseus are found toward an \av\ $\geq$ 5 mag. 

\begin{figure}
\figurenum{10}
\plotone{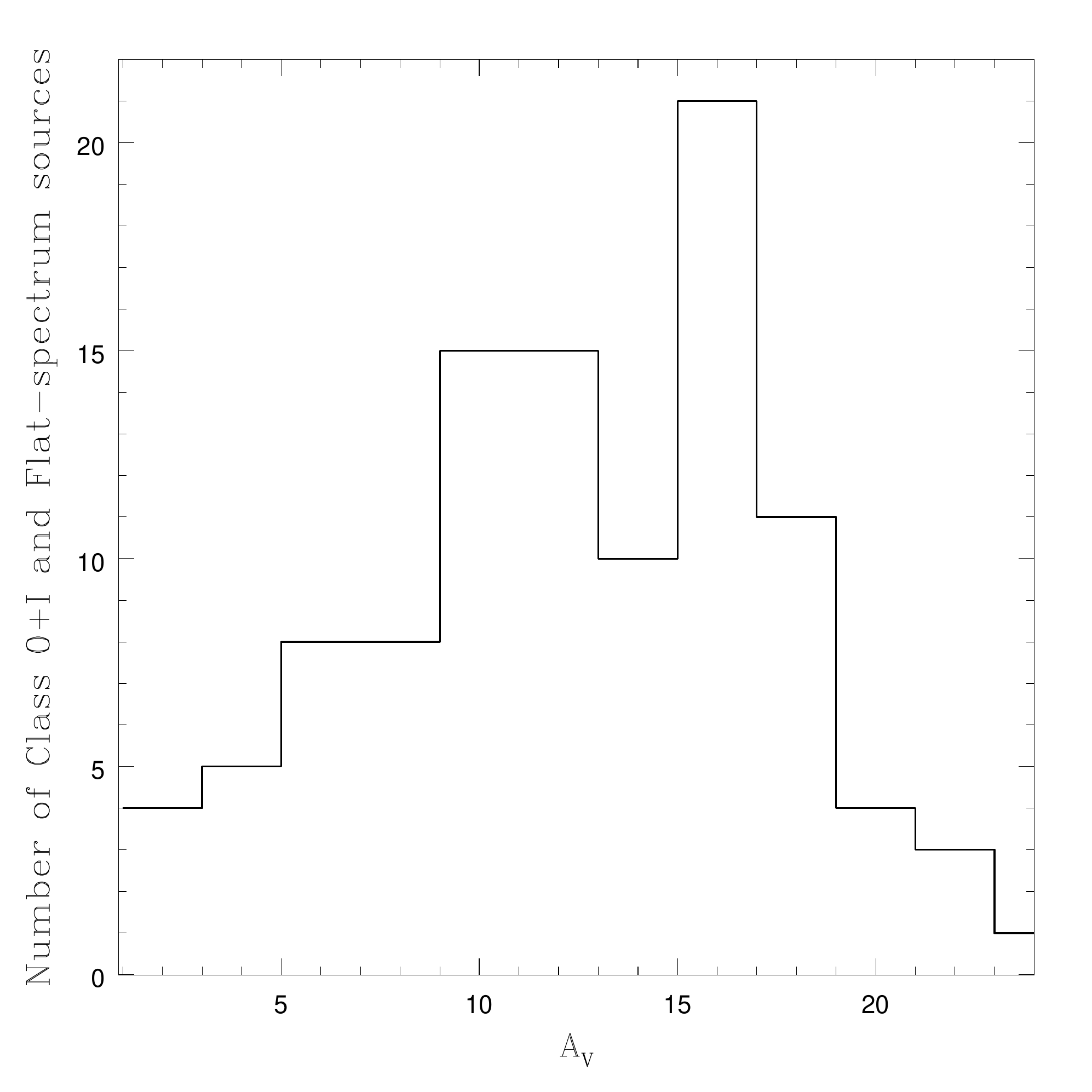}
\caption{Distribution of Class 0+I and Flat YSOs as a function of visual extinction at the source position using the background star extinctions averaged over 270\arcsec. 91\% of these sources are found toward \av\ $\geq$ 5 mag.}
\label{fig:ysohist}
\end{figure}

The distribution of the Class 0+I and Flat YSOs, those sources that are least likely to have dispersed from their place of origin, as a function of \av\ is shown in Figure \ref{fig:ysohist}. 91\% of the Class 0+I and Flat sources are found toward \av\ $\geq$ 5 mag. Only 3 Class 0+I and 6 Flat YSOs are found at \av \ $<$ 5 mag.  \citet{hei15} report 75\% of 535 Class 0+I and Flat sources in the c2d plus Gould Belt sample (M. M. Dunham et al. 2015, in preparation), including the Perseus YSOs in this work, are found toward \av\ $>$ 8 mag. We find 85\% of the Perseus Class 0+I and Flat sources are found toward \av\ $>$ 8 mag. These results agree with studies identifying dense cores only in regions of high extinction in Perseus \citep{hei10, eno08, kir06, hat05} and support the idea of a threshold extinction near \av\ = 5 - 8 mag.

\section{Summary}

We present IRAC and MIPS data from 3.6 to 70 \micron\ of the Perseus molecular cloud and give a broad overview of the current star formation in the region. Perseus contains 369 YSOs and has a density of 5 YSOs per square parsec. Tables \ref{tab:348} - \ref{tab:rc} list updated extinction corrected photometry for the YSOs including both absorption and scattering cross sections. While Perseus has a large fraction of embedded Class 0+I sources (19\%), 63\%\ of the objects in the cloud are classified as Class II YSOs.

Perseus contains a variety of star forming environments. IC 348 and the eastern end of the cloud are shrouded in extended nebulosity at the longer {\it Spitzer} wavelengths. However, the IC 348 region contains fewer embedded sources and the majority of the Class II (73\%) and III (64\%) objects in Perseus, consistent with previous age estimates of 2 Myr \citep{lad06}. The NGC 1333 cluster, suggested to be $<$ 1 Myr \citep{wil04}, contains more than twice the number of Class 0+I and Flat sources than IC 348. The ratios of Class II to Class 0+I YSOs (9.5 in IC 348 and 2.1 in NGC 1333) are also consistent with the relative age estimates of the clusters. 

We find that 56\% of all the YSOs and 91\% of the Class 0+I and Flat spectrum sources in Perseus are in regions where the \av\ $\geq$ 5 mag, consistent with an extinction threshold for early stage star formation suggested by other authors \citep[e.g.,][]{eno08, sad14, hei15}. However, the mid-cloud region is nearly devoid of star formation activity despite regions where the visual extinction ranges from \av\ = 5 - 20 mag. 

\acknowledgements
Support of this work, part of the {\it Spitzer} Legacy Science Program, was provided by NASA through contracts 1224608, 1230782, and 1230779 issued by the Jet Propulsion Laboratory, California Institute of Technology, under NASA contract 1407. K. E. Y. and C. H. Y. acknowledge support of a Louisiana Space Consortium Research Enhancement Award through NASA EPSCoR grant number NNX10AI40H. M. M. D. acknowledges support from the Submillimeter Array through an SMA postdoctoral fellowship. N. J. E. acknowledges support from a grant from the National Science Foundation, AST-1109116. We would like to thank the anonymous referee for comments that improved the focus and clarity of this work.

{\it Facility:} \facility{Spitzer}

\clearpage
\LongTables
\begin{deluxetable}{llcccccccc}
\tabletypesize{\scriptsize}
\tablecolumns{10}
\tablecaption{YSOs in the IC 348 Region \label{tab:348}}
\tablewidth{0pt}
\tablehead{
	\colhead{E09}  & 
	\colhead{Source Name}  & 
	\colhead{$\alpha$} &
	\colhead{Class} & 
	\colhead{IRAC\tablenotemark{a}} &
	\colhead{IRAC} &
	\colhead{IRAC} &
	\colhead{IRAC} &
	\colhead{MIPS\tablenotemark{a}} &
	\colhead{MIPS} \\
	\colhead{Index} &
	\colhead{(SSTc2d $+$)} &
	\colhead{} &
	\colhead{} &
	\colhead{3.6 \micron} &
	\colhead{4.5 \micron} &
	\colhead{5.8 \micron} &
	\colhead{8.0 \micron} &
	\colhead{24 \micron} &
	\colhead{70 \micron} 
}
\tablenotetext{a}{{\it Spitzer} Flux Densities in mJy}
\startdata
337	&	J034313.70+320045.2	&	-1.15	&	II	&	16	$\pm$	1.1	&	19	$\pm$	1.2	&	16	$\pm$	1.0	&	21	$\pm$	1.3	&	26	$\pm$	2.7	&	$\cdots$			\\
340	&	J034323.57+321225.9	&	-1.52	&	II	&	4.9	$\pm$	0.36	&	4.0	$\pm$	0.31	&	3.2	$\pm$	0.28	&	3.0	$\pm$	0.28	&	2.8	$\pm$	0.37	&	$\cdots$			\\
341	&	J034325.48+315516.5	&	-1.33	&	II	&	15	$\pm$	1.1	&	11	$\pm$	0.86	&	7.8	$\pm$	0.70	&	5.9	$\pm$	0.69	&	17	$\pm$	1.9	&	$\cdots$			\\
342	&	J034328.21+320159.1	&	-1.39	&	II	&	63	$\pm$	4.3	&	65	$\pm$	4.4	&	46	$\pm$	3.6	&	46	$\pm$	3.7	&	45	$\pm$	4.7	&	$\cdots$			\\
343	&	J034328.46+320505.8	&	-1.19	&	II	&	8.2	$\pm$	0.60	&	7.6	$\pm$	0.54	&	6.3	$\pm$	0.48	&	7.6	$\pm$	0.54	&	11	$\pm$	1.2	&	$\cdots$			\\
344	&	J034329.43+315219.5	&	0.41	&	0+I	&	7.7	$\pm$	0.53	&	13	$\pm$	0.71	&	20	$\pm$	1.0	&	43	$\pm$	2.1	&	180	$\pm$	20	&	260	$\pm$	33	\\
345	&	J034336.02+315009.0	&	-0.18	&	Flat	&	3.7	$\pm$	0.26	&	4.4	$\pm$	0.28	&	5.2	$\pm$	0.31	&	9.5	$\pm$	0.47	&	30	$\pm$	3.1	&	$\cdots$			\\
347	&	J034344.63+320817.8	&	-0.93	&	II	&	22	$\pm$	1.6	&	17	$\pm$	1.4	&	12	$\pm$	1.2	&	12	$\pm$	1.2	&	110	$\pm$	12	&	170	$\pm$	44	\\
348	&	J034345.17+320358.6	&	0.11	&	Flat	&	300	$\pm$	22	&	190	$\pm$	19	&	250	$\pm$	18	&	140	$\pm$	17	&	850	$\pm$	90	&	940	$\pm$	100	\\
349	&	J034348.83+321551.5	&	-1.07	&	II	&	10	$\pm$	0.79	&	9.9	$\pm$	0.71	&	10	$\pm$	0.69	&	13	$\pm$	0.8	&	15	$\pm$	1.6	&	$\cdots$			\\
350	&	J034350.96+320324.7	&	1.39	&	0+I	&	0.49	$\pm$	0.04	&	1.5	$\pm$	0.12	&	1.9	$\pm$	0.18	&	3.4	$\pm$	0.2	&	62	$\pm$	6.6	&	$\cdots$			\\
351	&	J034351.02+320308.1	&	-0.29	&	Flat	&	9.8	$\pm$	0.85	&	5.4	$\pm$	0.85	&	17	$\pm$	1.2	&	2.9	$\pm$	0.24	&	52	$\pm$	5.5	&	2000	$\pm$	290	\\
353	&	J034355.24+315532.1	&	-0.99	&	II	&	52	$\pm$	3.7	&	43	$\pm$	3.1	&	31	$\pm$	2.6	&	25	$\pm$	2.5	&	100	$\pm$	10	&	$\cdots$			\\
354	&	J034355.28+320753.3	&	-1.09	&	II	&	5.0	$\pm$	0.38	&	4.8	$\pm$	0.34	&	4.6	$\pm$	0.34	&	5.6	$\pm$	0.37	&	5.9	$\pm$	0.66	&	$\cdots$			\\
355	&	J034356.03+320213.3	&	-1.20	&	II	&	210	$\pm$	14	&	200	$\pm$	14	&	170	$\pm$	11	&	190	$\pm$	13	&	210	$\pm$	22	&	$\cdots$			\\
357	&	J034356.84+320304.7	&	1.83	&	0+I	&	0.07	$\pm$	0	&	0.27	$\pm$	0.08	&	0.66	$\pm$	0.08	&	0.79	$\pm$	0.16	&	13	$\pm$	1.4	&	3400	$\pm$	460	\\
358	&	J034357.23+320133.7	&	-1.28	&	II	&	7.8	$\pm$	0.53	&	7.4	$\pm$	0.51	&	5.9	$\pm$	0.43	&	6.2	$\pm$	0.46	&	6.5	$\pm$	0.75	&	$\cdots$			\\
362	&	J034358.57+321727.5	&	-1.10	&	II	&	33	$\pm$	2.6	&	31	$\pm$	2.2	&	25	$\pm$	1.9	&	31	$\pm$	2.2	&	52	$\pm$	5.4	&	$\cdots$			\\
363	&	J034358.92+321127.2	&	-0.84	&	II	&	46	$\pm$	3.5	&	45	$\pm$	3.2	&	38	$\pm$	2.8	&	50	$\pm$	3.3	&	170	$\pm$	17	&	$\cdots$			\\
364	&	J034359.10+321421.2	&	-1.28	&	II	&	20	$\pm$	1.5	&	21	$\pm$	1.4	&	16	$\pm$	1.2	&	19	$\pm$	1.3	&	27	$\pm$	3.0	&	$\cdots$			\\
366	&	J034359.65+320154.1	&	-0.90	&	II	&	1200	$\pm$	84	&	1300	$\pm$	97	&	1300	$\pm$	74	&	1800	$\pm$	110	&	1500	$\pm$	160	&	1700	$\pm$	190	\\
367	&	J034359.88+320441.4	&	-1.10	&	II	&	3.9	$\pm$	0.28	&	3.2	$\pm$	0.25	&	2.8	$\pm$	0.23	&	3.2	$\pm$	0.26	&	7	$\pm$	0.79	&	$\cdots$			\\
368	&	J034400.48+320432.7	&	-0.82	&	II	&	4.3	$\pm$	0.29	&	4.0	$\pm$	0.27	&	3.7	$\pm$	0.25	&	4.4	$\pm$	0.29	&	9.2	$\pm$	1.1	&	$\cdots$			\\
369	&	J034401.57+322358.9	&	-1.21	&	II	&	6.3	$\pm$	0.47	&	5.6	$\pm$	0.41	&	4.5	$\pm$	0.39	&	5.4	$\pm$	0.40	&	8.1	$\pm$	0.88	&	$\cdots$			\\
370	&	J034402.40+320204.9	&	1.47	&	0+I	&	1.1	$\pm$	0.07	&	2.6	$\pm$	0.2	&	4.8	$\pm$	0.34	&	8.4	$\pm$	0.59	&	130	$\pm$	14	&	870	$\pm$	120	\\
371	&	J034402.63+320159.5	&	0.92	&	0+I	&	1.2	$\pm$	0.08	&	3.8	$\pm$	0.22	&	5.1	$\pm$	0.25	&	8.0	$\pm$	0.39	&	63	$\pm$	7.5	&	$\cdots$			\\
372	&	J034402.92+315227.7	&	-1.31	&	II	&	5.5	$\pm$	0.39	&	4.7	$\pm$	0.35	&	4.3	$\pm$	0.32	&	4.7	$\pm$	0.35	&	4.4	$\pm$	0.51	&	$\cdots$			\\
373	&	J034405.78+320001.1	&	-0.92	&	II	&	5.7	$\pm$	0.40	&	5.2	$\pm$	0.34	&	4.7	$\pm$	0.32	&	5.7	$\pm$	0.36	&	9.9	$\pm$	1.1	&	$\cdots$			\\
374	&	J034405.78+320028.5	&	-0.64	&	II	&	55	$\pm$	3.8	&	65	$\pm$	4.0	&	68	$\pm$	3.8	&	68	$\pm$	3.7	&	31	$\pm$	3.3	&	$\cdots$			\\
375	&	J034406.01+321532.1	&	-1.16	&	II	&	3.0	$\pm$	0.21	&	2.7	$\pm$	0.21	&	2.7	$\pm$	0.20	&	2.8	$\pm$	0.22	&	2.9	$\pm$	0.38	&	$\cdots$			\\
377	&	J034406.80+320754.0	&	-1.15	&	II	&	13	$\pm$	0.95	&	11	$\pm$	0.82	&	9.5	$\pm$	0.77	&	11	$\pm$	0.8	&	15	$\pm$	1.6	&	$\cdots$			\\
378	&	J034409.16+320709.3	&	-2.36	&	III	&	98	$\pm$	7.2	&	66	$\pm$	5.9	&	46	$\pm$	5.4	&	37	$\pm$	5.3	&	10	$\pm$	1.1	&	$\cdots$			\\
379	&	J034409.20+320237.8	&	0.61	&	0+I	&	17	$\pm$	1.2	&	34	$\pm$	1.8	&	36	$\pm$	1.9	&	63	$\pm$	3.2	&	320	$\pm$	35	&	280	$\pm$	59	\\
380	&	J034410.13+320404.5	&	-1.11	&	II	&	10	$\pm$	0.72	&	9.3	$\pm$	0.69	&	8.3	$\pm$	0.64	&	9.0	$\pm$	0.73	&	16	$\pm$	1.9	&	$\cdots$			\\
381	&	J034411.63+320313.1	&	-1.39	&	II	&	92	$\pm$	6.5	&	68	$\pm$	5.6	&	68	$\pm$	5.0	&	70	$\pm$	5.3	&	50	$\pm$	5.4	&	$\cdots$			\\
382	&	J034412.98+320135.5	&	0.29	&	Flat	&	220	$\pm$	14	&	590	$\pm$	40	&	620	$\pm$	32	&	790	$\pm$	47	&	1800	$\pm$	200	&	4700	$\pm$	510	\\
383	&	J034415.84+315936.7	&	-1.03	&	II	&	4.0	$\pm$	0.27	&	3.4	$\pm$	0.26	&	3.1	$\pm$	0.24	&	4.1	$\pm$	0.28	&	7.6	$\pm$	0.84	&	$\cdots$			\\
384	&	J034418.10+321053.5	&	-1.03	&	II	&	50	$\pm$	3.6	&	41	$\pm$	2.9	&	32	$\pm$	2.3	&	35	$\pm$	2.5	&	39	$\pm$	4.2	&	$\cdots$			\\
385	&	J034418.17+320457.0	&	-0.92	&	II	&	130	$\pm$	8.9	&	83	$\pm$	7.4	&	93	$\pm$	6.9	&	79	$\pm$	6.8	&	580	$\pm$	61	&	$\cdots$			\\
386	&	J034418.21+320959.3	&	-1.45	&	II	&	8.2	$\pm$	0.63	&	6.9	$\pm$	0.53	&	5.6	$\pm$	0.48	&	5.2	$\pm$	0.48	&	9.0	$\pm$	1.0	&	$\cdots$			\\
387	&	J034418.27+320732.5	&	-1.49	&	II	&	3.8	$\pm$	0.27	&	2.8	$\pm$	0.24	&	2.1	$\pm$	0.23	&	2.2	$\pm$	0.24	&	1.9	$\pm$	0.42	&	$\cdots$			\\
388	&	J034418.59+321253.1	&	-1.01	&	II	&	37	$\pm$	2.8	&	37	$\pm$	2.5	&	35	$\pm$	2.4	&	36	$\pm$	2.5	&	36	$\pm$	3.9	&	$\cdots$			\\
389	&	J034419.14+320931.3	&	-2.38	&	III	&	32	$\pm$	2.5	&	23	$\pm$	2.0	&	16	$\pm$	1.8	&	9.8	$\pm$	1.7	&	3.8	$\pm$	0.51	&	$\cdots$			\\
391	&	J034419.25+320734.7	&	-1.30	&	II	&	16	$\pm$	1.1	&	14	$\pm$	1.1	&	13	$\pm$	1.0	&	14	$\pm$	1.1	&	8.3	$\pm$	0.94	&	$\cdots$			\\
392	&	J034420.18+320856.5	&	-1.66	&	III	&	12	$\pm$	0.90	&	11	$\pm$	0.78	&	9.3	$\pm$	0.73	&	9.8	$\pm$	0.79	&	6.0	$\pm$	0.89	&	$\cdots$			\\
393	&	J034421.23+320114.5	&	-1.39	&	II	&	7.3	$\pm$	0.51	&	5.3	$\pm$	0.42	&	3.9	$\pm$	0.34	&	4.0	$\pm$	0.36	&	8.9	$\pm$	0.97	&	$\cdots$			\\
394	&	J034421.29+321237.2	&	-1.48	&	II	&	5.0	$\pm$	0.39	&	4.8	$\pm$	0.34	&	4.0	$\pm$	0.32	&	3.5	$\pm$	0.32	&	3.2	$\pm$	0.42	&	$\cdots$			\\
395	&	J034421.31+321156.3	&	-1.35	&	II	&	14	$\pm$	1.1	&	12	$\pm$	0.88	&	11	$\pm$	0.86	&	14	$\pm$	0.97	&	15	$\pm$	1.6	&	$\cdots$			\\
396	&	J034421.35+315932.6	&	0.52	&	0+I	&	44	$\pm$	3.0	&	67	$\pm$	4.7	&	79	$\pm$	5.3	&	93	$\pm$	6.5	&	260	$\pm$	28	&	610	$\pm$	88	\\
397	&	J034421.57+321509.7	&	-1.73	&	III	&	6.3	$\pm$	0.47	&	4.7	$\pm$	0.38	&	3.2	$\pm$	0.35	&	2.2	$\pm$	0.33	&	5.0	$\pm$	0.66	&	$\cdots$			\\
398	&	J034421.62+321037.7	&	-1.09	&	II	&	75	$\pm$	5.4	&	67	$\pm$	4.8	&	53	$\pm$	4.2	&	50	$\pm$	4.2	&	68	$\pm$	7.2	&	$\cdots$			\\
399	&	J034422.29+320542.8	&	-0.89	&	II	&	41	$\pm$	2.8	&	30	$\pm$	2.5	&	28	$\pm$	2.3	&	40	$\pm$	2.7	&	130	$\pm$	13	&	$\cdots$			\\
400	&	J034422.34+321200.7	&	-1.13	&	II	&	20	$\pm$	1.5	&	20	$\pm$	1.4	&	21	$\pm$	1.4	&	26	$\pm$	1.7	&	26	$\pm$	2.7	&	$\cdots$			\\
401	&	J034422.58+320153.6	&	-1.05	&	II	&	20	$\pm$	1.4	&	13	$\pm$	1.2	&	10	$\pm$	1.1	&	8.7	$\pm$	1.1	&	73	$\pm$	7.6	&	$\cdots$			\\
402	&	J034422.70+320142.3	&	-1.68	&	III	&	4.6	$\pm$	0.32	&	3.4	$\pm$	0.27	&	2.7	$\pm$	0.24	&	2.6	$\pm$	0.26	&	7.8	$\pm$	1.8	&	$\cdots$			\\
403	&	J034423.65+320152.7	&	-1.36	&	II	&	13	$\pm$	0.91	&	11	$\pm$	0.8	&	9.7	$\pm$	0.71	&	11	$\pm$	0.82	&	10	$\pm$	1.2	&	$\cdots$			\\
404	&	J034424.46+320143.7	&	-1.25	&	II	&	2.6	$\pm$	0.19	&	2.7	$\pm$	0.18	&	2.2	$\pm$	0.17	&	2.4	$\pm$	0.19	&	4.2	$\pm$	1.0	&	$\cdots$			\\
405	&	J034424.84+321348.4	&	1.65	&	0+I	&	0.27	$\pm$	0.02	&	0.32	$\pm$	0.02	&	0.33	$\pm$	0.07	&	0.99	$\pm$	0.21	&	36	$\pm$	3.7	&	$\cdots$			\\
406	&	J034425.32+321012.7	&	-1.42	&	II	&	14	$\pm$	1.1	&	12	$\pm$	0.93	&	11	$\pm$	0.88	&	14	$\pm$	1.0	&	12	$\pm$	3.9	&	$\cdots$			\\
407	&	J034425.52+321131.2	&	-0.79	&	II	&	63	$\pm$	4.8	&	54	$\pm$	4.2	&	46	$\pm$	3.6	&	67	$\pm$	4.4	&	150	$\pm$	16	&	$\cdots$			\\
408	&	J034425.55+320617.1	&	-1.37	&	II	&	22	$\pm$	1.5	&	16	$\pm$	0.13	&	1.3	$\pm$	1.2	&	14	$\pm$	1.4	&	30	$\pm$	3.4	&	$\cdots$			\\
409	&	J034425.71+321549.2	&	-1.06	&	II	&	3.1	$\pm$	0.23	&	2.7	$\pm$	0.20	&	2.4	$\pm$	0.2	&	2.7	$\pm$	0.23	&	5.9	$\pm$	0.70	&	$\cdots$			\\
410	&	J034426.04+320430.4	&	-1.15	&	II	&	570	$\pm$	39	&	520	$\pm$	39	&	450	$\pm$	33	&	490	$\pm$	40	&	540	$\pm$	58	&	$\cdots$			\\
411	&	J034426.70+320820.3	&	-1.21	&	II	&	47	$\pm$	3.6	&	39	$\pm$	2.9	&	28	$\pm$	2.4	&	31	$\pm$	2.7	&	59	$\pm$	6.4	&	$\cdots$			\\
412	&	J034427.22+322028.8	&	-1.39	&	II	&	4.5	$\pm$	0.34	&	3.9	$\pm$	0.29	&	3.2	$\pm$	0.27	&	3.4	$\pm$	0.27	&	4.3	$\pm$	0.51	&	$\cdots$			\\
413	&	J034427.26+321420.9	&	-1.17	&	II	&	14	$\pm$	1.1	&	13	$\pm$	0.92	&	11	$\pm$	0.83	&	13	$\pm$	0.92	&	23	$\pm$	2.4	&	$\cdots$			\\
414	&	J034427.26+321037.3	&	-1.60	&	II	&	9.7	$\pm$	0.72	&	8.3	$\pm$	0.62	&	6.1	$\pm$	0.6	&	6.4	$\pm$	0.61	&	14	$\pm$	3.5	&	$\cdots$			\\
416	&	J034428.14+321600.2	&	-2.35	&	III	&	14	$\pm$	1.1	&	10	$\pm$	0.85	&	7.3	$\pm$	0.75	&	4.7	$\pm$	0.72	&	1.8	$\pm$	0.32	&	$\cdots$			\\
417	&	J034428.51+315954.0	&	-1.51	&	II	&	23	$\pm$	1.6	&	19	$\pm$	1.5	&	15	$\pm$	1.3	&	16	$\pm$	1.3	&	16	$\pm$	1.7	&	$\cdots$			\\
418	&	J034428.95+320137.9	&	-1.00	&	II	&	18	$\pm$	1.3	&	17	$\pm$	1.2	&	14	$\pm$	0.99	&	17	$\pm$	1.2	&	41	$\pm$	4.4	&	$\cdots$			\\
419	&	J034429.23+320115.7	&	-1.33	&	II	&	11	$\pm$	0.76	&	8.0	$\pm$	0.64	&	5.9	$\pm$	0.55	&	5.8	$\pm$	0.57	&	17	$\pm$	1.8	&	$\cdots$			\\
420	&	J034429.74+321039.8	&	-1.08	&	II	&	53	$\pm$	4.0	&	47	$\pm$	3.4	&	37	$\pm$	3.0	&	45	$\pm$	3.3	&	160	$\pm$	17	&	$\cdots$			\\
421	&	J034429.80+320054.6	&	-0.93	&	II	&	7.8	$\pm$	0.54	&	6.9	$\pm$	0.50	&	6.8	$\pm$	0.48	&	8.2	$\pm$	0.54	&	9.9	$\pm$	0.11	&	$\cdots$			\\
422	&	J034429.98+321922.6	&	-2.19	&	III	&	14	$\pm$	1.1	&	11	$\pm$	0.87	&	7.0	$\pm$	0.78	&	4.2	$\pm$	0.71	&	2.2	$\pm$	0.32	&	$\cdots$			\\
423	&	J034430.14+320118.2	&	-1.83	&	III	&	10	$\pm$	0.71	&	7.2	$\pm$	0.59	&	5.7	$\pm$	0.51	&	5.3	$\pm$	0.54	&	3.9	$\pm$	0.53	&	$\cdots$			\\
424	&	J034430.30+321135.2	&	0.05	&	Flat	&	3.3	$\pm$	0.24	&	4.0	$\pm$	0.27	&	6.2	$\pm$	0.41	&	8.4	$\pm$	0.66	&	20	$\pm$	2.5	&	$\cdots$			\\
425	&	J034430.84+320955.7	&	-1.46	&	II	&	51	$\pm$	3.7	&	35	$\pm$	3.1	&	41	$\pm$	3.3	&	74	$\pm$	4.8	&	26	$\pm$	6.8	&	$\cdots$			\\
426	&	J034431.13+321848.5	&	-0.92	&	II	&	8.6	$\pm$	0.64	&	8.5	$\pm$	0.59	&	7.8	$\pm$	0.53	&	9.8	$\pm$	0.61	&	18	$\pm$	1.9	&	$\cdots$			\\
427	&	J034431.19+320558.9	&	-0.59	&	II	&	13	$\pm$	0.88	&	13	$\pm$	0.94	&	13	$\pm$	0.9	&	13	$\pm$	0.91	&	31	$\pm$	3.3	&	$\cdots$			\\
428	&	J034431.37+320014.2	&	-1.28	&	II	&	69	$\pm$	5.3	&	63	$\pm$	4.7	&	53	$\pm$	3.9	&	67	$\pm$	4.5	&	77	$\pm$	8.1	&	$\cdots$			\\
429	&	J034431.55+320844.9	&	-2.36	&	III	&	42	$\pm$	3.2	&	29	$\pm$	2.5	&	20	$\pm$	2.3	&	13	$\pm$	2.2	&	12	$\pm$	3.8	&	$\cdots$			\\
430	&	J034432.04+321143.7	&	-0.62	&	II	&	200	$\pm$	16	&	150	$\pm$	13	&	330	$\pm$	22	&	1000	$\pm$	66	&	430	$\pm$	45	&	$\cdots$			\\
431	&	J034432.59+320855.7	&	-1.93	&	III	&	23	$\pm$	1.7	&	16	$\pm$	1.4	&	14	$\pm$	1.2	&	13	$\pm$	1.3	&	15	$\pm$	4.0	&	$\cdots$			\\
432	&	J034433.22+321257.4	&	-1.35	&	II	&	6.4	$\pm$	0.47	&	4.9	$\pm$	0.38	&	4.1	$\pm$	0.35	&	4.6	$\pm$	0.44	&	7.0	$\pm$	0.89	&	$\cdots$			\\
433	&	J034433.79+315830.2	&	-1.25	&	II	&	14	$\pm$	1.0	&	12	$\pm$	0.88	&	9.2	$\pm$	0.77	&	9.7	$\pm$	0.80	&	24	$\pm$	2.5	&	$\cdots$			\\
434	&	J034434.05+320657.0	&	-0.81	&	II	&	4.6	$\pm$	0.32	&	4.3	$\pm$	0.32	&	4.5	$\pm$	0.32	&	5.5	$\pm$	0.40	&	5.6	$\pm$	1.3	&	$\cdots$			\\
435	&	J034434.14+321635.7	&	-1.20	&	II	&	6.0	$\pm$	0.45	&	5.3	$\pm$	0.39	&	4.5	$\pm$	0.37	&	5.0	$\pm$	0.39	&	7.9	$\pm$	0.94	&	$\cdots$			\\
436	&	J034434.21+320946.3	&	-2.02	&	III	&	620	$\pm$	54	&	380	$\pm$	39	&	300	$\pm$	35	&	170	$\pm$	32	&	350	$\pm$	53	&	$\cdots$			\\
437	&	J034434.29+321240.7	&	-1.34	&	II	&	29	$\pm$	2.2	&	29	$\pm$	2.0	&	23	$\pm$	1.6	&	27	$\pm$	1.8	&	30	$\pm$	3.2	&	$\cdots$			\\
438	&	J034434.69+321600.1	&	-0.43	&	II	&	17	$\pm$	1.3	&	19	$\pm$	1.2	&	18	$\pm$	1.1	&	21	$\pm$	1.3	&	150	$\pm$	15	&	$\cdots$			\\
439	&	J034434.71+321554.4	&	-1.98	&	III	&	12	$\pm$	0.96	&	9.3	$\pm$	0.74	&	6.3	$\pm$	0.64	&	3.6	$\pm$	0.62	&	7.2	$\pm$	1.4	&	$\cdots$			\\
440	&	J034434.81+315655.2	&	-1.19	&	II	&	13	$\pm$	0.94	&	10	$\pm$	0.82	&	7.4	$\pm$	0.73	&	5.9	$\pm$	0.71	&	23	$\pm$	2.5	&	$\cdots$			\\
441	&	J034434.99+321531.1	&	-1.29	&	II	&	10	$\pm$	0.78	&	9.8	$\pm$	0.69	&	9.1	$\pm$	0.67	&	12	$\pm$	0.77	&	8.6	$\pm$	0.94	&	$\cdots$			\\
443	&	J034435.38+320736.2	&	-1.20	&	II	&	37	$\pm$	2.7	&	38	$\pm$	2.7	&	30	$\pm$	2.2	&	46	$\pm$	2.9	&	55	$\pm$	5.8	&	$\cdots$			\\
444	&	J034435.38+321004.6	&	-1.57	&	II	&	430	$\pm$	37	&	350	$\pm$	32	&	290	$\pm$	26	&	300	$\pm$	27	&	270	$\pm$	30	&	$\cdots$			\\
445	&	J034435.47+320856.3	&	-1.13	&	II	&	12	$\pm$	0.90	&	10	$\pm$	0.78	&	9.3	$\pm$	0.79	&	11	$\pm$	1.2	&	50	$\pm$	11	&	$\cdots$			\\
446	&	J034435.69+320303.5	&	-1.06	&	II	&	21	$\pm$	1.5	&	19	$\pm$	1.4	&	18	$\pm$	1.3	&	22	$\pm$	0.14	&	27	$\pm$	3.1	&	$\cdots$			\\
448	&	J034436.96+320645.2	&	-1.93	&	III	&	230	$\pm$	15	&	160	$\pm$	14	&	130	$\pm$	13	&	98	$\pm$	12	&	47	$\pm$	5.3	&	$\cdots$			\\
449	&	J034437.40+321224.2	&	-0.93	&	II	&	21	$\pm$	1.5	&	16	$\pm$	1.3	&	12	$\pm$	1.1	&	15	$\pm$	1.3	&	77	$\pm$	8.2	&	$\cdots$			\\
450	&	J034437.88+320804.1	&	-1.22	&	II	&	97	$\pm$	6.8	&	93	$\pm$	6.5	&	68	$\pm$	5.4	&	87	$\pm$	6.2	&	100	$\pm$	11	&	$\cdots$			\\
451	&	J034437.98+320329.7	&	-1.24	&	II	&	70	$\pm$	4.8	&	60	$\pm$	4.6	&	51	$\pm$	4.0	&	72	$\pm$	4.7	&	75	$\pm$	8.1	&	$\cdots$			\\
452	&	J034438.01+321137.0	&	-1.49	&	II	&	8.2	$\pm$	0.6	&	6.5	$\pm$	0.51	&	6.0	$\pm$	0.5	&	7.0	$\pm$	0.61	&	10	$\pm$	2.9	&	$\cdots$			\\
453	&	J034438.46+320735.7	&	-1.13	&	II	&	55	$\pm$	3.8	&	44	$\pm$	3.5	&	44	$\pm$	3.3	&	73	$\pm$	4.6	&	88	$\pm$	9.4	&	$\cdots$			\\
454	&	J034438.54+320800.6	&	-0.97	&	II	&	36	$\pm$	2.5	&	29	$\pm$	2.3	&	23	$\pm$	2.0	&	35	$\pm$	2.6	&	120	$\pm$	13	&	$\cdots$			\\
455	&	J034438.97+320319.7	&	-1.25	&	II	&	4.1	$\pm$	0.29	&	3.4	$\pm$	0.26	&	2.9	$\pm$	0.25	&	3.8	$\pm$	0.37	&	5.9	$\pm$	1.6	&	$\cdots$			\\
456	&	J034439.19+322008.9	&	-1.80	&	III	&	13	$\pm$	0.95	&	10	$\pm$	0.80	&	8.6	$\pm$	0.75	&	8.3	$\pm$	0.81	&	6.2	$\pm$	1.8	&	$\cdots$			\\
457	&	J034439.21+320944.8	&	-1.44	&	II	&	22	$\pm$	1.6	&	19	$\pm$	1.4	&	16	$\pm$	1.4	&	18	$\pm$	1.6	&	52	$\pm$	15	&	$\cdots$			\\
458	&	J034439.80+321804.0	&	-1.40	&	II	&	27	$\pm$	2.0	&	21	$\pm$	1.7	&	17	$\pm$	1.5	&	18	$\pm$	1.5	&	24	$\pm$	2.5	&	$\cdots$			\\
459	&	J034441.74+321202.2	&	-1.07	&	II	&	15	$\pm$	1.1	&	11	$\pm$	0.88	&	8.6	$\pm$	0.81	&	8.9	$\pm$	0.85	&	56	$\pm$	5.9	&	$\cdots$			\\
460	&	J034442.04+320859.9	&	-2.01	&	III	&	41	$\pm$	3.2	&	34	$\pm$	2.6	&	25	$\pm$	2.4	&	26	$\pm$	2.6	&	17	$\pm$	4.0	&	$\cdots$			\\
461	&	J034442.15+320902.1	&	-0.52	&	II	&	26	$\pm$	2.2	&	35	$\pm$	2.2	&	36	$\pm$	2.2	&	39	$\pm$	2.4	&	74	$\pm$	7.8	&	$\cdots$			\\
462	&	J034442.58+321002.5	&	-1.29	&	II	&	14	$\pm$	1.1	&	14	$\pm$	0.99	&	13	$\pm$	0.91	&	14	$\pm$	1.0	&	16	$\pm$	2.5	&	$\cdots$			\\
463	&	J034442.76+320833.7	&	-1.35	&	II	&	14	$\pm$	1.0	&	12	$\pm$	0.91	&	12	$\pm$	0.89	&	13	$\pm$	1.1	&	15	$\pm$	3.4	&	$\cdots$			\\
464	&	J034443.03+321559.6	&	-0.65	&	II	&	5.4	$\pm$	0.39	&	5.1	$\pm$	0.36	&	5.0	$\pm$	0.37	&	7.9	$\pm$	0.49	&	20	$\pm$	2.1	&	$\cdots$			\\
466	&	J034443.32+320131.5	&	0.61	&	0+I	&	260	$\pm$	19	&	400	$\pm$	27	&	480	$\pm$	25	&	740	$\pm$	38	&	2300	$\pm$	250	&	$\cdots$			\\
467	&	J034443.53+320742.7	&	-1.97	&	III	&	36	$\pm$	2.6	&	23	$\pm$	2.0	&	16	$\pm$	1.8	&	11	$\pm$	1.9	&	11	$\pm$	1.5	&	$\cdots$			\\
468	&	J034443.78+321030.4	&	-0.87	&	II	&	24	$\pm$	1.7	&	23	$\pm$	1.6	&	24	$\pm$	1.6	&	43	$\pm$	2.4	&	74	$\pm$	7.9	&	$\cdots$			\\
469	&	J034443.96+320136.2	&	0.74	&	0+I	&	3.9	$\pm$	0.30	&	8.5	$\pm$	0.89	&	7.0	$\pm$	0.41	&	$\cdots$			&	170	$\pm$	23	&	11000	$\pm$	1200	\\
470	&	J034444.59+320812.5	&	-0.91	&	II	&	33	$\pm$	2.3	&	32	$\pm$	2.2	&	27	$\pm$	2.0	&	39	$\pm$	2.4	&	30	$\pm$	4.0	&	$\cdots$			\\
471	&	J034444.72+320402.5	&	-1.18	&	II	&	130	$\pm$	9.1	&	120	$\pm$	8.8	&	100	$\pm$	7.7	&	120	$\pm$	8.4	&	170	$\pm$	18	&	$\cdots$			\\
472	&	J034445.20+320119.6	&	-0.58	&	II	&	18	$\pm$	1.2	&	17	$\pm$	1.2	&	19	$\pm$	1.1	&	23	$\pm$	1.3	&	35	$\pm$	3.8	&	$\cdots$			\\
473	&	J034450.35+315236.0	&	-1.12	&	II	&	4.0	$\pm$	0.28	&	3.7	$\pm$	0.26	&	3.0	$\pm$	0.24	&	3.0	$\pm$	0.24	&	5.2	$\pm$	0.57	&	$\cdots$			\\
475	&	J034452.05+315825.2	&	-1.27	&	II	&	16	$\pm$	1.1	&	1.4	$\pm$	0.99	&	11	$\pm$	0.89	&	9.9	$\pm$	0.87	&	15	$\pm$	1.5	&	$\cdots$			\\
476	&	J034455.62+320919.8	&	-2.23	&	III	&	33	$\pm$	2.3	&	22	$\pm$	1.9	&	14	$\pm$	1.6	&	9.2	$\pm$	1.5	&	10	$\pm$	2.3	&	$\cdots$			\\
477	&	J034456.14+320915.2	&	-1.12	&	II	&	110	$\pm$	7.8	&	79	$\pm$	6.7	&	65	$\pm$	6.0	&	53	$\pm$	5.8	&	230	$\pm$	25	&	$\cdots$			\\
478	&	J034456.84+315411.4	&	-1.35	&	II	&	3.0	$\pm$	0.2	&	2.2	$\pm$	0.18	&	1.7	$\pm$	0.16	&	1.2	$\pm$	0.17	&	3.3	$\pm$	0.39	&	$\cdots$			\\
479	&	J034456.90+322035.6	&	-0.72	&	II	&	2.9	$\pm$	0.21	&	3.1	$\pm$	0.19	&	2.7	$\pm$	0.18	&	2.5	$\pm$	0.21	&	6.3	$\pm$	0.80	&	$\cdots$			\\
480	&	J034457.72+320741.7	&	-1.10	&	II	&	3.1	$\pm$	0.21	&	2.3	$\pm$	0.18	&	1.7	$\pm$	0.17	&	1.4	$\pm$	0.2	&	6.9	$\pm$	0.8	&	$\cdots$			\\
481	&	J034457.85+320401.5	&	-1.82	&	III	&	4.9	$\pm$	0.33	&	3.8	$\pm$	0.30	&	3.0	$\pm$	0.28	&	3.2	$\pm$	0.29	&	1.6	$\pm$	0.35	&	$\cdots$			\\
482	&	J034458.55+315827.1	&	-1.19	&	II	&	2.3	$\pm$	0.16	&	2.0	$\pm$	0.15	&	1.8	$\pm$	0.13	&	2.2	$\pm$	0.15	&	1.7	$\pm$	0.30	&	$\cdots$			\\
483	&	J034459.85+321331.9	&	-1.68	&	III	&	6.6	$\pm$	0.46	&	5.2	$\pm$	0.41	&	4.4	$\pm$	0.39	&	4.4	$\pm$	0.41	&	2.3	$\pm$	0.37	&	$\cdots$			\\
484	&	J034500.45+320320.1	&	-1.78	&	III	&	5.3	$\pm$	0.37	&	4.1	$\pm$	0.32	&	3.2	$\pm$	0.28	&	2.5	$\pm$	0.28	&	2.1	$\pm$	0.38	&	$\cdots$			\\
485	&	J034501.43+320501.7	&	-2.31	&	III	&	44	$\pm$	3.0	&	30	$\pm$	2.6	&	20	$\pm$	2.4	&	14	$\pm$	2.3	&	4.4	$\pm$	0.51	&	$\cdots$			\\
486	&	J034504.66+321501.1	&	-2.10	&	III	&	10	$\pm$	0.75	&	7.3	$\pm$	0.60	&	5.3	$\pm$	0.53	&	4.4	$\pm$	0.52	&	2.5	$\pm$	0.82	&	$\cdots$			\\
487	&	J034507.64+321027.9	&	-2.48	&	III	&	40	$\pm$	3.5	&	34	$\pm$	2.6	&	24	$\pm$	2.3	&	15	$\pm$	2.1	&	3.9	$\pm$	0.49	&	$\cdots$			\\
488	&	J034513.08+322005.3	&	-1.48	&	II	&	2.6	$\pm$	0.2	&	2.4	$\pm$	0.17	&	1.9	$\pm$	0.15	&	2.1	$\pm$	0.17	&	1.7	$\pm$	0.27	&	$\cdots$			\\
490	&	J034513.82+321210.0	&	0.43	&	0+I	&	2.9	$\pm$	0.2	&	4.7	$\pm$	0.27	&	8.1	$\pm$	0.42	&	16	$\pm$	0.76	&	69	$\pm$	7.4	&	290	$\pm$	38	\\
491	&	J034516.34+320620.1	&	-0.71	&	II	&	240	$\pm$	17	&	280	$\pm$	18	&	260	$\pm$	16	&	370	$\pm$	24	&	450	$\pm$	48	&	470	$\pm$	59	\\
492	&	J034517.82+321205.8	&	-1.52	&	II	&	6.3	$\pm$	0.47	&	5.5	$\pm$	0.41	&	4.2	$\pm$	0.38	&	4.0	$\pm$	0.36	&	4.3	$\pm$	0.53	&	$\cdots$			\\
493	&	J034520.46+320634.5	&	-1.13	&	II	&	130	$\pm$	8.9	&	120	$\pm$	8.4	&	91	$\pm$	7.2	&	110	$\pm$	7.9	&	230	$\pm$	24	&	370	$\pm$	42	\\
494	&	J034525.15+320930.3	&	-0.83	&	II	&	41	$\pm$	2.7	&	35	$\pm$	2.6	&	31	$\pm$	2.4	&	47	$\pm$	2.9	&	120	$\pm$	12	&	91	$\pm$	16	\\
495	&	J034529.72+315919.7	&	-1.29	&	II	&	4.6	$\pm$	0.31	&	3.6	$\pm$	0.27	&	3.2	$\pm$	0.26	&	3.3	$\pm$	0.27	&	5.1	$\pm$	0.58	&	$\cdots$			\\
497	&	J034535.64+315954.4	&	-1.32	&	II	&	12	$\pm$	0.81	&	9.4	$\pm$	0.71	&	8.1	$\pm$	0.65	&	9.9	$\pm$	0.71	&	14	$\pm$	1.5	&	$\cdots$			

\enddata

\end{deluxetable}

\clearpage
\begin{deluxetable}{llcccccccc}
\tabletypesize{\scriptsize}
\tablecolumns{10}
\tablecaption{YSOs in the NGC 1333 Region \label{tab:1333}}
\tablewidth{0pt}
\tablehead{
	\colhead{E09}  & 
	\colhead{Source Name}  & 
	\colhead{$\alpha$} &
	\colhead{Class} & 
	\colhead{IRAC\tablenotemark{a}} &
	\colhead{IRAC} &
	\colhead{IRAC} &
	\colhead{IRAC} &
	\colhead{MIPS\tablenotemark{a}} &
	\colhead{MIPS} \\
	\colhead{Index} &
	\colhead{(SSTc2d $+$)} &
	\colhead{} &
	\colhead{} &
	\colhead{3.6 \micron} &
	\colhead{4.5 \micron} &
	\colhead{5.8 \micron} &
	\colhead{8.0 \micron} &
	\colhead{24 \micron} &
	\colhead{70 \micron} 
}
\tablenotetext{a}{{\it Spitzer} Flux Densities in mJy}
\startdata
136	&	J032832.56+311105.1	&	0.51	&	0+I	&	1.4	$\pm$	0.13	&	3.1	$\pm$	0.23	&	3.3	$\pm$	0.23	&	5.3	$\pm$	0.37	&	63	$\pm$	6.7	&	63	$\pm$	6.7	\\
138	&	J032834.49+310051.1	&	0.31	&	0+I	&	6.6	$\pm$	0.46	&	16	$\pm$	0.86	&	31	$\pm$	1.5	&	45	$\pm$	2.2	&	69	$\pm$	7.4	&	160	$\pm$	19	\\
141	&	J032837.09+311330.8	&	1.93	&	0+I	&	43	$\pm$	3.3	&	120	$\pm$	7.7	&	340	$\pm$	17	&	950	$\pm$	54	&	7900	$\pm$	1400	&	55000	$\pm$	5800	\\
142	&	J032838.78+311806.6	&	-0.25	&	Flat	&	6.5	$\pm$	0.46	&	9.4	$\pm$	0.58	&	10	$\pm$	0.59	&	13	$\pm$	0.70	&	18	$\pm$	2.0	&	$\cdots$			\\
143	&	J032839.10+310601.7	&	1.55	&	0+I	&	0.19	$\pm$	0	&	0.36	$\pm$	0.03	&	0.55	$\pm$	0.05	&	1.2	$\pm$	0.09	&	28	$\pm$	3.0	&	96	$\pm$	15	\\
144	&	J032839.71+311731.9	&	0.61	&	0+I	&	19	$\pm$	1.4	&	31	$\pm$	1.8	&	41	$\pm$	2.2	&	55	$\pm$	2.8	&	220	$\pm$	22	&	370	$\pm$	41	\\
145	&	J032840.63+311756.5	&	0.96	&	0+I	&	0.04	$\pm$	0.01	&	0.26	$\pm$	0.02	&	0.62	$\pm$	0.05	&	1.2	$\pm$	0.08	&	6.0	$\pm$	0.69	&	$\cdots$			\\
147	&	J032843.24+311042.7	&	-1.86	&	III	&	23	$\pm$	1.6	&	16	$\pm$	1.3	&	13	$\pm$	1.1	&	11	$\pm$	1.1	&	5.8	$\pm$	0.66	&	$\cdots$			\\
148	&	J032843.28+311732.9	&	0.24	&	Flat	&	260	$\pm$	20	&	470	$\pm$	28	&	620	$\pm$	38	&	1000	$\pm$	61	&	2200	$\pm$	240	&	2100	$\pm$	240	\\
149	&	J032844.09+312052.7	&	-1.34	&	II	&	6.4	$\pm$	0.47	&	5.2	$\pm$	0.41	&	4.7	$\pm$	0.37	&	5.2	$\pm$	0.39	&	4.8	$\pm$	0.53	&	$\cdots$			\\
151	&	J032846.21+311638.4	&	-1.84	&	III	&	51	$\pm$	4.0	&	43	$\pm$	3.3	&	38	$\pm$	3.1	&	36	$\pm$	3.0	&	15	$\pm$	1.6	&	$\cdots$			\\
152	&	J032847.65+312406.0	&	-1.32	&	II	&	35	$\pm$	2.7	&	29	$\pm$	2.1	&	25	$\pm$	1.9	&	20	$\pm$	1.9	&	21	$\pm$	2.1	&	$\cdots$			\\
153	&	J032847.84+311655.1	&	-1.10	&	II	&	39	$\pm$	3.0	&	46	$\pm$	3.1	&	38	$\pm$	2.5	&	43	$\pm$	2.7	&	53	$\pm$	5.6	&	$\cdots$			\\
154	&	J032848.77+311608.8	&	0.20	&	Flat	&	2.3	$\pm$	0.17	&	3.1	$\pm$	0.19	&	4.3	$\pm$	0.24	&	7.2	$\pm$	0.37	&	22	$\pm$	2.3	&	$\cdots$			\\
156	&	J032850.62+304244.7	&	-0.95	&	II	&	220	$\pm$	19	&	200	$\pm$	16	&	180	$\pm$	15	&	270	$\pm$	22	&	390	$\pm$	40	&	$\cdots$			\\
157	&	J032851.08+311632.4	&	-1.48	&	II	&	6.3	$\pm$	0.49	&	5.6	$\pm$	0.42	&	5.1	$\pm$	0.4	&	5.8	$\pm$	0.42	&	4.0	$\pm$	0.49	&	$\cdots$			\\
158	&	J032851.20+311954.8	&	-1.00	&	II	&	78	$\pm$	5.8	&	72	$\pm$	5.2	&	69	$\pm$	4.9	&	110	$\pm$	6.4	&	150	$\pm$	16	&	$\cdots$			\\
159	&	J032851.26+311739.3	&	0.60	&	0+I	&	0.96	$\pm$	0.08	&	1.6	$\pm$	0.10	&	2.2	$\pm$	0.01	&	6.5	$\pm$	0.42	&	29	$\pm$	3.0	&	$\cdots$			\\
160	&	J032852.15+311547.1	&	-1.34	&	II	&	8.8	$\pm$	0.65	&	7.3	$\pm$	0.56	&	6.6	$\pm$	0.53	&	8.1	$\pm$	0.58	&	5.3	$\pm$	0.62	&	$\cdots$			\\
162	&	J032852.17+312245.3	&	-1.45	&	II	&	27	$\pm$	2.1	&	22	$\pm$	1.8	&	19	$\pm$	1.6	&	22	$\pm$	1.6	&	25	$\pm$	2.6	&	$\cdots$			\\
163	&	J032852.92+311626.4	&	-1.49	&	II	&	4.8	$\pm$	0.37	&	4.2	$\pm$	0.32	&	3.8	$\pm$	0.3	&	3.9	$\pm$	0.31	&	3.2	$\pm$	0.46	&	$\cdots$			\\
164	&	J032853.96+311809.3	&	-1.29	&	II	&	69	$\pm$	5.3	&	60	$\pm$	4.2	&	51	$\pm$	3.6	&	70	$\pm$	4.4	&	130	$\pm$	13	&	350	$\pm$	47	\\
165	&	J032854.09+311654.2	&	-1.07	&	II	&	15	$\pm$	1.1	&	12	$\pm$	0.91	&	11	$\pm$	0.9	&	16	$\pm$	1.0	&	23	$\pm$	2.7	&	$\cdots$			\\
166	&	J032854.63+311651.1	&	-1.08	&	II	&	94	$\pm$	7.1	&	85	$\pm$	6.2	&	80	$\pm$	5.5	&	110	$\pm$	6.6	&	140	$\pm$	14	&	$\cdots$			\\
167	&	J032855.08+311628.7	&	-0.76	&	II	&	140	$\pm$	9.9	&	120	$\pm$	8.8	&	110	$\pm$	7.7	&	130	$\pm$	8.9	&	250	$\pm$	26	&	$\cdots$			\\
168	&	J032855.55+311436.7	&	2.39	&	0+I	&	1.4	$\pm$	0.12	&	17	$\pm$	1.2	&	29	$\pm$	1.8	&	41	$\pm$	2.8	&	$\cdots$			&	140000	$\pm$	15000	\\
169	&	J032856.12+311908.4	&	-0.51	&	II	&	8.5	$\pm$	0.62	&	25	$\pm$	1.6	&	34	$\pm$	1.7	&	34	$\pm$	1.7	&	24	$\pm$	2.6	&	$\cdots$			\\
170	&	J032856.32+312227.9	&	-0.40	&	II	&	23	$\pm$	1.7	&	24	$\pm$	1.6	&	29	$\pm$	1.8	&	41	$\pm$	2.2	&	100	$\pm$	11	&	560	$\pm$	72	\\
171	&	J032856.60+310737.0	&	0.01	&	Flat	&	1.1	$\pm$	0.09	&	3.4	$\pm$	0.22	&	2.5	$\pm$	0.17	&	0.87	$\pm$	0.11	&	14	$\pm$	1.4	&	110	$\pm$	14	\\
172	&	J032856.65+311835.5	&	-1.14	&	II	&	150	$\pm$	11	&	130	$\pm$	10	&	90	$\pm$	7.5	&	100	$\pm$	8.1	&	280	$\pm$	29	&	650	$\pm$	76	\\
173	&	J032856.97+311622.3	&	-0.74	&	II	&	24	$\pm$	1.9	&	20	$\pm$	1.6	&	17	$\pm$	1.4	&	26	$\pm$	1.7	&	210	$\pm$	22	&	$\cdots$			\\
174	&	J032857.18+311534.6	&	-0.71	&	II	&	5.1	$\pm$	0.40	&	4.8	$\pm$	0.35	&	5.3	$\pm$	0.37	&	8.2	$\pm$	0.45	&	16	$\pm$	1.7	&	$\cdots$			\\
175	&	J032857.21+311419.1	&	1.40	&	0+I	&	46	$\pm$	3.2	&	130	$\pm$	8.1	&	330	$\pm$	17	&	500	$\pm$	30	&	5000	$\pm$	530	&	31000	$\pm$	3300	\\
177	&	J032857.70+311948.1	&	-0.96	&	II	&	19	$\pm$	1.5	&	19	$\pm$	1.3	&	18	$\pm$	1.2	&	26	$\pm$	1.6	&	41	$\pm$	4.4	&	$\cdots$			\\
178	&	J032858.11+311803.7	&	-2.37	&	III	&	12	$\pm$	0.92	&	8.7	$\pm$	0.74	&	6.3	$\pm$	0.66	&	4.7	$\pm$	0.64	&	1.3	$\pm$	0.28	&	$\cdots$			\\
179	&	J032858.26+312209.2	&	-0.99	&	II	&	5.3	$\pm$	0.38	&	4.9	$\pm$	0.38	&	4.5	$\pm$	0.34	&	5.7	$\pm$	1.1	&	44	$\pm$	12	&	$\cdots$			\\
180	&	J032858.27+312202.0	&	-1.07	&	II	&	13	$\pm$	1.0	&	14	$\pm$	0.9	&	13	$\pm$	0.87	&	17	$\pm$	0.99	&	19	$\pm$	4.1	&	$\cdots$			\\
181	&	J032858.43+312217.5	&	0.82	&	0+I	&	57	$\pm$	4.3	&	110	$\pm$	11	&	230	$\pm$	12	&	340	$\pm$	21	&	990	$\pm$	110	&	950	$\pm$	110	\\
182	&	J032859.23+312032.5	&	-0.02	&	Flat	&	0.90	$\pm$	0.10	&	2.0	$\pm$	0.19	&	1.7	$\pm$	0.17	&	1.7	$\pm$	0.24	&	13	$\pm$	2.3	&	$\cdots$			\\
183	&	J032859.32+311548.7	&	-0.08	&	Flat	&	160	$\pm$	11	&	160	$\pm$	12	&	150	$\pm$	11	&	250	$\pm$	14	&	970	$\pm$	100	&	$\cdots$			\\
184	&	J032859.56+312146.7	&	-1.11	&	II	&	120	$\pm$	9.0	&	120	$\pm$	8.1	&	96	$\pm$	6.8	&	110	$\pm$	7.4	&	12	$\pm$	13	&	$\cdots$			\\
185	&	J032900.55+311200.8	&	1.96	&	0+I	&	$\cdots$			&	0.12	$\pm$	0	&	0.14	$\pm$	0.04	&	$\cdots$			&	17	$\pm$	1.8	&	1300	$\pm$	180	\\
186	&	J032901.56+312020.6	&	2.30	&	0+I	&	770	$\pm$	66	&	1500	$\pm$	110	&	2200	$\pm$	270	&	4100	$\pm$	250	&	$\cdots$			&	24000	$\pm$	2600	\\
188	&	J032902.81+312217.2	&	-1.32	&	II	&	9.5	$\pm$	0.72	&	9.1	$\pm$	0.61	&	8.5	$\pm$	0.57	&	9.6	$\pm$	0.62	&	7.9	$\pm$	2.1	&	$\cdots$			\\
189	&	J032903.15+312238.0	&	-0.76	&	II	&	35	$\pm$	2.5	&	38	$\pm$	2.5	&	35	$\pm$	2.3	&	52	$\pm$	3.0	&	100	$\pm$	11	&	$\cdots$			\\
190	&	J032903.22+312545.1	&	-0.81	&	II	&	2.6	$\pm$	0.19	&	2.3	$\pm$	0.18	&	2.1	$\pm$	0.17	&	2.4	$\pm$	0.19	&	9.2	$\pm$	1.0	&	$\cdots$			\\
191	&	J032903.33+312314.6	&	1.00	&	0+I	&	6.5	$\pm$	0.47	&	10	$\pm$	0.59	&	24	$\pm$	1.2	&	61	$\pm$	2.9	&	160	$\pm$	17	&	$\cdots$			\\
192	&	J032903.78+311603.8	&	0.85	&	0+I	&	1300	$\pm$	100	&	480	$\pm$	110	&	3600	$\pm$	420	&	7500	$\pm$	430	&	$\cdots$			&	130000	$\pm$	13000	\\
194	&	J032903.87+312148.6	&	-1.06	&	II	&	260	$\pm$	20	&	230	$\pm$	18	&	210	$\pm$	15	&	280	$\pm$	19	&	320	$\pm$	34	&	$\cdots$			\\
195	&	J032904.06+311446.5	&	1.32	&	0+I	&	0.63	$\pm$	0.05	&	1.5	$\pm$	0.12	&	1.7	$\pm$	0.10	&	1.8	$\pm$	0.12	&	83	$\pm$	8.8	&	$\cdots$			\\
198	&	J032904.68+311659.0	&	-0.46	&	II	&	11	$\pm$	0.83	&	12	$\pm$	0.78	&	12	$\pm$	0.79	&	20	$\pm$	1.1	&	41	$\pm$	4.5	&	$\cdots$			\\
199	&	J032904.73+311134.9	&	-0.62	&	II	&	14	$\pm$	0.94	&	17	$\pm$	1.1	&	16	$\pm$	0.89	&	19	$\pm$	1.1	&	17	$\pm$	1.8	&	$\cdots$			\\
200	&	J032904.95+312038.4	&	0.06	&	Flat	&	13	$\pm$	0.95	&	16	$\pm$	1.0	&	19	$\pm$	1.1	&	28	$\pm$	1.5	&	110	$\pm$	12	&	$\cdots$			\\
201	&	J032905.18+312036.9	&	-0.98	&	II	&	2.9	$\pm$	0.27	&	4.3	$\pm$	0.27	&	5.3	$\pm$	0.34	&	7.8	$\pm$	0.42	&	38	$\pm$	6.0	&	$\cdots$			\\
202	&	J032905.78+311639.6	&	-1.08	&	II	&	470	$\pm$	37	&	460	$\pm$	33	&	410	$\pm$	27	&	500	$\pm$	30	&	450	$\pm$	48	&	$\cdots$			\\
204	&	J032906.33+311346.4	&	-1.14	&	II	&	48	$\pm$	3.4	&	47	$\pm$	3.1	&	43	$\pm$	2.5	&	52	$\pm$	3.1	&	55	$\pm$	5.9	&	$\cdots$			\\
205	&	J032907.78+312157.3	&	2.30	&	0+I	&	1000	$\pm$	110	&	2400	$\pm$	230	&	3900	$\pm$	360	&	3400	$\pm$	330	&	$\cdots$			&	50000	$\pm$	5300	\\
206	&	J032907.96+312251.4	&	-1.16	&	II	&	130	$\pm$	10	&	120	$\pm$	8.6	&	120	$\pm$	8.0	&	160	$\pm$	13	&	110	$\pm$	31	&	$\cdots$			\\
207	&	J032908.97+312256.1	&	-0.21	&	Flat	&	43	$\pm$	3.2	&	54	$\pm$	3.5	&	56	$\pm$	4.6	&	75	$\pm$	16	&	140	$\pm$	31	&	$\cdots$			\\
208	&	J032909.10+312305.5	&	-0.33	&	II	&	22	$\pm$	1.6	&	23	$\pm$	1.5	&	23	$\pm$	2.9	&	70	$\pm$	16	&	240	$\pm$	46	&	$\cdots$			\\
209	&	J032909.10+312128.7	&	1.05	&	0+I	&	39	$\pm$	3.3	&	110	$\pm$	6.9	&	170	$\pm$	9.8	&	170	$\pm$	13	&	430	$\pm$	63	&	$\cdots$			\\
210	&	J032909.34+312104.1	&	-0.51	&	II	&	5.9	$\pm$	0.44	&	4.7	$\pm$	0.38	&	5.1	$\pm$	0.50	&	8.6	$\pm$	0.76	&	41	$\pm$	5.6	&	$\cdots$			\\
211	&	J032909.40+311413.8	&	-0.87	&	II	&	2.9	$\pm$	0.20	&	7.3	$\pm$	0.51	&	8.9	$\pm$	0.45	&	6.0	$\pm$	0.35	&	3.6	$\pm$	0.50	&	$\cdots$			\\
212	&	J032909.49+312720.9	&	-1.10	&	II	&	5.2	$\pm$	0.39	&	4.7	$\pm$	0.35	&	4.2	$\pm$	0.32	&	6.0	$\pm$	0.39	&	7.4	$\pm$	0.85	&	$\cdots$			\\
213	&	J032909.65+312256.3	&	-1.60	&	II	&	96	$\pm$	6.9	&	68	$\pm$	5.7	&	56	$\pm$	6.0	&	87	$\pm$	16	&	120	$\pm$	29	&	$\cdots$			\\
214	&	J032910.47+312334.7	&	-1.41	&	II	&	10	$\pm$	0.72	&	8.9	$\pm$	0.62	&	8.3	$\pm$	0.60	&	9.5	$\pm$	0.87	&	13	$\pm$	3.3	&	$\cdots$			\\
215	&	J032910.49+311331.0	&	2.49	&	0+I	&	$\cdots$			&	0.11	$\pm$	0.01	&	0.17	$\pm$	0.03	&	$\cdots$			&	37	$\pm$	3.9	&	34000	$\pm$	3600	\\
216	&	J032910.65+311340.0	&	0.42	&	0+I	&	0.07	$\pm$	0.01	&	1.3	$\pm$	0.08	&	0.74	$\pm$	0.05	&	0.37	$\pm$	0.06	&	5.5	$\pm$	0.92	&	$\cdots$			\\
217	&	J032910.68+311820.6	&	1.84	&	0+I	&	2.6	$\pm$	0.26	&	9.0	$\pm$	0.89	&	13	$\pm$	1.1	&	14	$\pm$	1.0	&	890	$\pm$	94	&	21000	$\pm$	2300	\\
218	&	J032910.84+311642.6	&	-0.32	&	II	&	8.0	$\pm$	0.65	&	10	$\pm$	0.69	&	13	$\pm$	0.76	&	22	$\pm$	1.2	&	28	$\pm$	2.9	&	$\cdots$			\\
220	&	J032911.26+311831.4	&	1.79	&	0+I	&	1.0	$\pm$	0.11	&	5.2	$\pm$	0.35	&	6.4	$\pm$	0.42	&	5.4	$\pm$	0.36	&	600	$\pm$	63	&	$\cdots$			\\
221	&	J032911.89+312127.0	&	0.32	&	0+I	&	23	$\pm$	1.7	&	34	$\pm$	2.0	&	41	$\pm$	2.5	&	51	$\pm$	4.0	&	60	$\pm$	20	&	$\cdots$			\\
222	&	J032912.06+311305.4	&	0.85	&	0+I	&	0.04	$\pm$	0.01	&	1.3	$\pm$	0.09	&	0.80	$\pm$	0.05	&	0.39	$\pm$	0.11	&	21	$\pm$	2.9	&	12000	$\pm$	1300	\\
223	&	J032912.06+311301.7	&	0.07	&	Flat	&	0.98	$\pm$	0.07	&	17	$\pm$	0.91	&	6.2	$\pm$	0.32	&	4.5	$\pm$	0.28	&	44	$\pm$	4.6	&	$\cdots$			\\
224	&	J032912.97+311814.3	&	1.56	&	0+I	&	66	$\pm$	5.2	&	240	$\pm$	13	&	340	$\pm$	18	&	420	$\pm$	24	&	850	$\pm$	89	&	4400	$\pm$	470	\\
225	&	J032913.14+312252.8	&	-1.33	&	II	&	140	$\pm$	11	&	140	$\pm$	9.3	&	120	$\pm$	8.1	&	120	$\pm$	8.4	&	77	$\pm$	8.3	&	$\cdots$			\\
226	&	J032913.54+311358.2	&	2.50	&	0+I	&	0.04	$\pm$	0	&	$\cdots$			&	0.32	$\pm$	0.05	&	$\cdots$			&	32	$\pm$	3.5	&	4000	$\pm$	550	\\
227	&	J032914.40+311444.1	&	-0.70	&	II	&	4.4	$\pm$	0.35	&	16	$\pm$	1.5	&	8.1	$\pm$	0.65	&	6.7	$\pm$	0.50	&	12	$\pm$	1.3	&	$\cdots$			\\
228	&	J032916.61+312349.4	&	-1.18	&	II	&	32	$\pm$	2.3	&	31	$\pm$	2.1	&	32	$\pm$	2.1	&	43	$\pm$	2.5	&	21	$\pm$	3.1	&	$\cdots$			\\
229	&	J032916.69+311618.2	&	-2.43	&	III	&	24	$\pm$	1.7	&	17	$\pm$	1.5	&	12	$\pm$	1.3	&	7.2	$\pm$	1.3	&	1.9	$\pm$	0.31	&	$\cdots$			\\
230	&	J032916.83+312325.1	&	-0.57	&	II	&	3.0	$\pm$	0.22	&	2.7	$\pm$	0.20	&	2.9	$\pm$	0.22	&	4.2	$\pm$	0.25	&	23	$\pm$	2.7	&	$\cdots$			\\
231	&	J032917.17+312746.5	&	1.69	&	0+I	&	0.6	$\pm$	0.06	&	1.6	$\pm$	0.16	&	1.5	$\pm$	0.13	&	1.1	$\pm$	0.14	&	130	$\pm$	14	&	2000	$\pm$	300	\\
233	&	J032917.68+312245.0	&	-0.94	&	II	&	450	$\pm$	34	&	430	$\pm$	31	&	490	$\pm$	33	&	520	$\pm$	36	&	580	$\pm$	61	&	$\cdots$			\\
234	&	J032917.78+311948.0	&	-1.23	&	II	&	5.2	$\pm$	0.38	&	4.5	$\pm$	0.33	&	3.9	$\pm$	0.31	&	4.4	$\pm$	0.33	&	5.5	$\pm$	0.61	&	$\cdots$			\\
235	&	J032918.26+312319.9	&	1.09	&	0+I	&	1.3	$\pm$	0.10	&	3.8	$\pm$	0.2	&	8.0	$\pm$	0.38	&	14	$\pm$	0.72	&	68	$\pm$	7.7	&	$\cdots$			\\
236	&	J032918.67+312017.7	&	0.16	&	Flat	&	1.6	$\pm$	0.12	&	1.8	$\pm$	0.11	&	1.8	$\pm$	0.13	&	3.6	$\pm$	0.21	&	25	$\pm$	2.7	&	$\cdots$			\\
237	&	J032918.74+312325.4	&	-0.66	&	II	&	52	$\pm$	4.0	&	56	$\pm$	3.9	&	67	$\pm$	3.9	&	120	$\pm$	7.0	&	210	$\pm$	23	&	$\cdots$			\\
238	&	J032920.06+312407.5	&	0.34	&	0+I	&	43	$\pm$	3.4	&	59	$\pm$	3.6	&	76	$\pm$	5.0	&	120	$\pm$	9.9	&	430	$\pm$	45	&	$\cdots$			\\
239	&	J032920.44+311834.2	&	-0.25	&	Flat	&	130	$\pm$	11	&	160	$\pm$	10	&	150	$\pm$	9.4	&	210	$\pm$	12	&	510	$\pm$	54	&	210	$\pm$	30	\\
240	&	J032921.57+312110.3	&	-1.24	&	II	&	12	$\pm$	0.88	&	1.1	$\pm$	0.80	&	9.2	$\pm$	0.75	&	11	$\pm$	0.80	&	16	$\pm$	1.8	&	$\cdots$			\\
241	&	J032921.87+311536.2	&	-1.34	&	II	&	120	$\pm$	8.4	&	120	$\pm$	8.3	&	87	$\pm$	6.8	&	93	$\pm$	7.2	&	95	$\pm$	9.9	&	$\cdots$			\\
242	&	J032923.17+312030.2	&	-1.03	&	II	&	19	$\pm$	1.5	&	18	$\pm$	1.3	&	15	$\pm$	1.2	&	20	$\pm$	1.3	&	31	$\pm$	3.3	&	$\cdots$			\\
243	&	J032923.25+312653.1	&	-1.17	&	II	&	8.5	$\pm$	0.65	&	8.5	$\pm$	0.59	&	7.7	$\pm$	0.55	&	8.9	$\pm$	0.59	&	7.6	$\pm$	0.84	&	$\cdots$			\\
245	&	J032924.09+311957.6	&	-0.13	&	Flat	&	3.0	$\pm$	0.22	&	2.7	$\pm$	0.19	&	2.7	$\pm$	0.20	&	4.6	$\pm$	0.26	&	77	$\pm$	8.2	&	$\cdots$			\\
246	&	J032925.93+312640.1	&	-0.93	&	II	&	100	$\pm$	8.7	&	120	$\pm$	8.2	&	140	$\pm$	8.4	&	210	$\pm$	13	&	150	$\pm$	16	&	$\cdots$			\\
247	&	J032926.81+312647.6	&	-1.81	&	III	&	53	$\pm$	4.0	&	35	$\pm$	3.3	&	25	$\pm$	2.9	&	19	$\pm$	2.8	&	20	$\pm$	2.1	&	$\cdots$			\\
249	&	J032929.27+311834.7	&	-0.98	&	II	&	18	$\pm$	1.4	&	12	$\pm$	1.1	&	9.1	$\pm$	0.93	&	8.5	$\pm$	0.93	&	110	$\pm$	11	&	290	$\pm$	34	\\
250	&	J032929.80+312102.6	&	-1.15	&	II	&	18	$\pm$	1.4	&	14	$\pm$	1.1	&	15	$\pm$	1.1	&	21	$\pm$	1.3	&	31	$\pm$	3.3	&	$\cdots$			\\
251	&	J032930.40+311903.3	&	-1.11	&	II	&	18	$\pm$	1.3	&	17	$\pm$	1.3	&	17	$\pm$	1.2	&	20	$\pm$	1.3	&	29	$\pm$	3.1	&	120	$\pm$	29	\\
252	&	J032932.57+312436.9	&	-1.22	&	II	&	27	$\pm$	2.0	&	24	$\pm$	1.7	&	18	$\pm$	1.5	&	19	$\pm$	1.5	&	21	$\pm$	2.2	&	$\cdots$			\\
253	&	J032932.88+312712.6	&	-1.58	&	II	&	6.3	$\pm$	0.48	&	5.2	$\pm$	0.41	&	4.4	$\pm$	0.37	&	4.2	$\pm$	0.38	&	2.6	$\pm$	0.31	&	$\cdots$			\\
254	&	J032937.73+312202.5	&	-0.97	&	II	&	3.7	$\pm$	0.26	&	3.0	$\pm$	0.23	&	2.6	$\pm$	0.22	&	3.2	$\pm$	0.24	&	7.9	$\pm$	0.87	&	$\cdots$			\\
255	&	J032944.16+311947.3	&	-0.76	&	II	&	1.8	$\pm$	0.13	&	2.4	$\pm$	0.16	&	2.4	$\pm$	0.15	&	2.6	$\pm$	0.15	&	2.1	$\pm$	0.28	&	$\cdots$			\\
258	&	J032954.04+312052.9	&	-0.87	&	II	&	34	$\pm$	2.6	&	34	$\pm$	2.3	&	31	$\pm$	2.2	&	39	$\pm$	2.4	&	120	$\pm$	13	&	200	$\pm$	22	

\enddata

\end{deluxetable}

\clearpage
\begin{deluxetable}{llcccccccc}
\tabletypesize{\scriptsize}
\tablecolumns{10}
\tablecaption{YSOs in the RC \label{tab:rc}}
\tablewidth{0pt}
\tablehead{
	\colhead{E09}  & 
	\colhead{Source Name}  & 
	\colhead{$\alpha$} &
	\colhead{Class} & 
	\colhead{IRAC\tablenotemark{a}} &
	\colhead{IRAC} &
	\colhead{IRAC} &
	\colhead{IRAC} &
	\colhead{MIPS\tablenotemark{a}} &
	\colhead{MIPS} \\
	\colhead{Index} &
	\colhead{(SSTc2d $+$)} &
	\colhead{} &
	\colhead{} &
	\colhead{3.6 \micron} &
	\colhead{4.5 \micron} &
	\colhead{5.8 \micron} &
	\colhead{8.0 \micron} &
	\colhead{24 \micron} &
	\colhead{70 \micron} 
}
\tablenotetext{a}{{\it Spitzer} Flux Densities in mJy}
\startdata
121	&	J032519.52+303424.2	&	-0.14	&	Flat	&	3.6	$\pm$	0.26	&	4.4	$\pm$	0.27	&	5.3	$\pm$	0.31	&	7.0	$\pm$	0.38	&	13	$\pm$	1.4	&	$\cdots$			\\
122	&	J032522.32+304513.9	&	2.21	&	0+I	&	0.47	$\pm$	0.06	&	4.4	$\pm$	0.39	&	6.6	$\pm$	0.46	&	14	$\pm$	0.75	&	550	$\pm$	59	&	13000	$\pm$	1400	\\
123	&	J032536.22+304515.7	&	1.51	&	0+I	&	0.69	$\pm$	0.26	&	3.8	$\pm$	0.5	&	5.1	$\pm$	0.5	&	6.6	$\pm$	0.67	&	220	$\pm$	33	&	$\cdots$			\\
124	&	J032536.49+304522.2	&	2.56	&	0+I	&	1.0	$\pm$	0.22	&	15	$\pm$	0.99	&	57	$\pm$	2.8	&	180	$\pm$	9.3	&	4500	$\pm$	480	&	25000	$\pm$	2600	\\
125	&	J032538.83+304406.2	&	2.05	&	0+I	&	4.3	$\pm$	0.39	&	16	$\pm$	1.6	&	20	$\pm$	1.5	&	26	$\pm$	2.1	&	2100	$\pm$	220	&	32000	$\pm$	3400	\\
126	&	J032539.12+304358.2	&	2.00	&	0+I	&	3.4	$\pm$	0.27	&	31	$\pm$	1.9	&	94	$\pm$	4.7	&	170	$\pm$	8.7	&	1400	$\pm$	160	&	$\cdots$			\\
127	&	J032637.47+301528.1	&	1.03	&	0+I	&	5.2	$\pm$	0.42	&	13	$\pm$	0.87	&	14	$\pm$	0.98	&	16	$\pm$	1.1	&	460	$\pm$	48	&	4400	$\pm$	600	\\
128	&	J032738.25+301358.6	&	-0.40	&	II	&	90	$\pm$	6.3	&	97	$\pm$	6.9	&	100	$\pm$	7.0	&	120	$\pm$	8.7	&	390	$\pm$	41	&	1100	$\pm$	160	\\
130	&	J032739.08+301303.1	&	2.46	&	0+I	&	0.83	$\pm$	0.07	&	12	$\pm$	0.93	&	25	$\pm$	1.4	&	32	$\pm$	1.6	&	2000	$\pm$	210	&	21000	$\pm$	2400	\\
131	&	J032741.47+302016.8	&	-1.06	&	II	&	66	$\pm$	4.6	&	55	$\pm$	4.2	&	49	$\pm$	3.8	&	59	$\pm$	4.1	&	120	$\pm$	12	&	160	$\pm$	19	\\
132	&	J032743.23+301228.9	&	2.23	&	0+I	&	1.1	$\pm$	0.10	&	7.3	$\pm$	0.50	&	15	$\pm$	0.83	&	37	$\pm$	1.7	&	880	$\pm$	92	&	7600	$\pm$	800	\\
133	&	J032747.67+301204.5	&	-0.25	&	Flat	&	590	$\pm$	46	&	690	$\pm$	50	&	1000	$\pm$	56	&	1700	$\pm$	89	&	2000	$\pm$	210	&	1800	$\pm$	200	\\
134	&	J032800.09+300847.0	&	-1.13	&	II	&	54	$\pm$	3.8	&	46	$\pm$	3.4	&	35	$\pm$	3.0	&	33	$\pm$	3.0	&	92	$\pm$	9.7	&	160	$\pm$	20	\\
135	&	J032800.39+300801.3	&	0.98	&	0+I	&	19	$\pm$	1.3	&	36	$\pm$	2.0	&	57	$\pm$	2.8	&	92	$\pm$	4.5	&	430	$\pm$	45	&	700	$\pm$	76	\\
137	&	J032834.49+310051.1	&	0.88	&	0+I	&	79	$\pm$	5.5	&	160	$\pm$	10	&	280	$\pm$	14	&	320	$\pm$	17	&	1400	$\pm$	140	&	2800	$\pm$	300	\\
139	&	J032834.94+305454.5	&	-0.06	&	Flat	&	2.7	$\pm$	0.19	&	3.5	$\pm$	0.22	&	4.7	$\pm$	0.25	&	6.3	$\pm$	0.33	&	14	$\pm$	1.4	&	$\cdots$			\\
140	&	J032835.03+302009.9	&	-0.20	&	Flat	&	2.2	$\pm$	0.14	&	1.6	$\pm$	0.13	&	1.3	$\pm$	0.12	&	1.8	$\pm$	0.14	&	48	$\pm$	5.1	&	140	$\pm$	18	\\
146	&	J032842.41+302953.2	&	-1.05	&	II	&	33	$\pm$	2.3	&	30	$\pm$	2.3	&	29	$\pm$	2.1	&	36	$\pm$	2.4	&	43	$\pm$	4.5	&	$\cdots$			\\
150	&	J032845.30+310541.9	&	1.09	&	0+I	&	1.7	$\pm$	0.13	&	3.6	$\pm$	0.26	&	3.4	$\pm$	0.24	&	3.6	$\pm$	0.25	&	240	$\pm$	26	&	1500	$\pm$	210	\\
155	&	J032850.62+304244.7	&	-0.36	&	II	&	13	$\pm$	0.9	&	13	$\pm$	0.89	&	12	$\pm$	0.8	&	16	$\pm$	0.95	&	71	$\pm$	7.5	&	76	$\pm$	1.0	\\
161	&	J032852.17+304505.5	&	-0.78	&	II	&	300	$\pm$	22	&	350	$\pm$	30	&	410	$\pm$	23	&	520	$\pm$	30	&	340	$\pm$	36	&	$\cdots$			\\
193	&	J032903.87+305629.8	&	-0.44	&	II	&	1.4	$\pm$	0.10	&	1.3	$\pm$	0.09	&	1.1	$\pm$	0.09	&	1.3	$\pm$	0.09	&	13	$\pm$	1.4	&	$\cdots$			\\
196	&	J032904.12+305612.8	&	-0.97	&	II	&	4.2	$\pm$	0.28	&	3.6	$\pm$	0.27	&	3.2	$\pm$	0.26	&	3.9	$\pm$	0.28	&	9.2	$\pm$	1.0	&	$\cdots$			\\
203	&	J032906.05+303039.2	&	0.64	&	0+I	&	0.89	$\pm$	0.06	&	1.3	$\pm$	0.08	&	2.3	$\pm$	0.14	&	5.3	$\pm$	0.26	&	24	$\pm$	2.5	&	110	$\pm$	19	\\
244	&	J032923.48+313329.5	&	1.53	&	0+I	&	1.1	$\pm$	0.11	&	3.1	$\pm$	0.23	&	3.2	$\pm$	0.31	&	4.0	$\pm$	0.28	&	220	$\pm$	23	&	1200	$\pm$	160	\\
248	&	J032928.88+305841.9	&	-1.05	&	II	&	5.0	$\pm$	0.35	&	4.6	$\pm$	0.33	&	4.7	$\pm$	0.33	&	7.5	$\pm$	0.43	&	14	$\pm$	1.5	&	$\cdots$			\\
256	&	J032946.00+310439.0	&	-0.89	&	II	&	33	$\pm$	2.3	&	29	$\pm$	2.1	&	25	$\pm$	2.0	&	25	$\pm$	2.0	&	42	$\pm$	4.4	&	96	$\pm$	14	\\
257	&	J032951.82+313906.0	&	3.40	&	0+I	&	$\cdots$			&	0.03	$\pm$	0.01	&	0.13	$\pm$	0.04	&	0.49	$\pm$	0.05	&	56	$\pm$	6.0	&	2000	$\pm$	290	\\
259	&	J033015.14+302349.4	&	1.58	&	0+I	&	10	$\pm$	0.76	&	26	$\pm$	1.3	&	28	$\pm$	1.4	&	59	$\pm$	2.9	&	2100	$\pm$	220	&	5700	$\pm$	610	\\
260	&	J033022.45+313240.5	&	0.28	&	Flat	&	1.7	$\pm$	0.12	&	2.2	$\pm$	0.13	&	2.8	$\pm$	0.17	&	5.7	$\pm$	0.28	&	21	$\pm$	2.2	&	$\cdots$			\\
261	&	J033024.08+311404.4	&	-1.20	&	II	&	5.3	$\pm$	0.37	&	4.2	$\pm$	0.35	&	3.9	$\pm$	0.32	&	5.2	$\pm$	0.36	&	5.5	$\pm$	0.61	&	$\cdots$			\\
262	&	J033025.97+310217.9	&	-2.44	&	III	&	75	$\pm$	5.3	&	48	$\pm$	4.4	&	32	$\pm$	3.9	&	20	$\pm$	3.8	&	4.8	$\pm$	0.53	&	$\cdots$			\\
263	&	J033027.14+302829.8	&	-0.16	&	Flat	&	22	$\pm$	1.4	&	20	$\pm$	1.3	&	14	$\pm$	1.1	&	12	$\pm$	1.1	&	500	$\pm$	52	&	950	$\pm$	100	\\
264	&	J033032.68+302626.6	&	2.66	&	0+I	&	0.09	$\pm$		&	0.27	$\pm$	0	&	0.24	$\pm$	0.05	&	0.21	$\pm$	0.04	&	36	$\pm$	3.7	&	320	$\pm$	44	\\
265	&	J033035.47+311558.6	&	-0.60	&	II	&	2.0	$\pm$	0.14	&	1.6	$\pm$	0.12	&	1.5	$\pm$	0.13	&	2.0	$\pm$	0.14	&	20	$\pm$	2.1	&	130	$\pm$	20	\\
266	&	J033035.92+303024.4	&	-1.21	&	II	&	720	$\pm$	56	&	560	$\pm$	47	&	520	$\pm$	42	&	550	$\pm$	44	&	600	$\pm$	64	&	180	$\pm$	20	\\
267	&	J033036.97+303127.7	&	-0.97	&	II	&	260	$\pm$	18	&	280	$\pm$	18	&	310	$\pm$	18	&	430	$\pm$	24	&	410	$\pm$	42	&	120	$\pm$	16	\\
268	&	J033038.20+303211.9	&	-0.02	&	Flat	&	1.6	$\pm$	0.11	&	1.3	$\pm$	0.10	&	1.1	$\pm$	0.10	&	0.92	$\pm$	0.09	&	64	$\pm$	6.8	&	250	$\pm$	30	\\
269	&	J033043.98+303247.0	&	-0.89	&	II	&	410	$\pm$	33	&	390	$\pm$	28	&	380	$\pm$	28	&	430	$\pm$	28	&	780	$\pm$	83	&	810	$\pm$	86	\\
270	&	J033052.51+305417.8	&	-0.86	&	II	&	49	$\pm$	3.5	&	47	$\pm$	3.2	&	44	$\pm$	3.0	&	71	$\pm$	4.0	&	120	$\pm$	13	&	160	$\pm$	21	\\
271	&	J033110.68+304940.6	&	-0.92	&	II	&	31	$\pm$	2.2	&	29	$\pm$	2.1	&	30	$\pm$	2.1	&	39	$\pm$	2.5	&	64	$\pm$	6.8	&	82	$\pm$	13	\\
272	&	J033114.70+304955.4	&	-0.07	&	Flat	&	55	$\pm$	3.9	&	67	$\pm$	4.2	&	81	$\pm$	4.6	&	110	$\pm$	5.5	&	260	$\pm$	28	&	230	$\pm$	26	\\
273	&	J033118.30+304939.5	&	-1.01	&	II	&	230	$\pm$	17	&	210	$\pm$	16	&	200	$\pm$	15	&	280	$\pm$	18	&	340	$\pm$	35	&	200	$\pm$	25	\\
274	&	J033120.11+304917.7	&	-0.89	&	II	&	9.8	$\pm$	0.68	&	8.8	$\pm$	0.64	&	8.3	$\pm$	0.61	&	9.9	$\pm$	0.66	&	16	$\pm$	1.7	&	$\cdots$			\\
275	&	J033120.98+304530.1	&	1.48	&	0+I	&	0.19	$\pm$		&	1.1	$\pm$	0.09	&	1.1	$\pm$	0.09	&	0.82	$\pm$	0.08	&	17	$\pm$	1.7	&	4000	$\pm$	540	\\
276	&	J033128.87+303053.3	&	-1.36	&	II	&	220	$\pm$	16	&	190	$\pm$	14	&	160	$\pm$	13	&	140	$\pm$	13	&	140	$\pm$	15	&	210	$\pm$	26	\\
277	&	J033131.17+304411.1	&	-1.46	&	II	&	1.6	$\pm$	0.22	&	5.3	$\pm$	0.48	&	2.3	$\pm$	0.23	&	2.4	$\pm$	0.24	&	1.9	$\pm$	0.30	&	$\cdots$			\\
278	&	J033142.40+310624.9	&	-1.48	&	II	&	45	$\pm$	3.2	&	35	$\pm$	2.7	&	26	$\pm$	2.3	&	24	$\pm$	2.4	&	29	$\pm$	3.1	&	$\cdots$			\\
279	&	J033217.96+304947.5	&	0.86	&	0+I	&	0.56	$\pm$	0.11	&	0.56	$\pm$	0.19	&	$\cdots$			&	1.7	$\pm$	0.14	&	15	$\pm$	1.6	&	1700	$\pm$	180	\\
280	&	J033229.17+310240.8	&	0.16	&	Flat	&	16	$\pm$	1.2	&	23	$\pm$	1.8	&	29	$\pm$	2.0	&	43	$\pm$	3.0	&	210	$\pm$	22	&	1100	$\pm$	160	\\
281	&	J033232.99+310221.7	&	-1.12	&	II	&	83	$\pm$	6.0	&	79	$\pm$	5.7	&	66	$\pm$	4.8	&	78	$\pm$	5.4	&	90	$\pm$	9.5	&	$\cdots$			\\
282	&	J033234.05+310055.8	&	-1.02	&	II	&	140	$\pm$	9.7	&	120	$\pm$	9.0	&	100	$\pm$	7.7	&	120	$\pm$	8.8	&	340	$\pm$	36	&	300	$\pm$	36	\\
284	&	J033241.70+311046.3	&	-0.78	&	II	&	37	$\pm$	2.7	&	33	$\pm$	2.3	&	29	$\pm$	2.2	&	55	$\pm$	3.8	&	130	$\pm$	14	&	97	$\pm$	13	\\
285	&	J033247.20+305916.3	&	-1.27	&	II	&	530	$\pm$	35	&	480	$\pm$	34	&	440	$\pm$	28	&	500	$\pm$	31	&	360	$\pm$	39	&	210	$\pm$	28	\\
286	&	J033257.84+310608.3	&	0.33	&	0+I	&	1.2	$\pm$	0.08	&	1.3	$\pm$	0.08	&	1.4	$\pm$	0.10	&	1.6	$\pm$	0.1	&	17	$\pm$	1.7	&	83	$\pm$	14	\\
287	&	J033306.41+310804.6	&	0.17	&	Flat	&	10	$\pm$	0.73	&	13	$\pm$	0.79	&	14	$\pm$	0.83	&	28	$\pm$	1.3	&	110	$\pm$	11	&	45	$\pm$	9.9	\\
288	&	J033309.56+310531.2	&	0.70	&	0+I	&	0.76	$\pm$	0.05	&	0.67	$\pm$	0.05	&	0.71	$\pm$	0.07	&	2.0	$\pm$	0.12	&	89	$\pm$	9.3	&	100	$\pm$	14	\\
289	&	J033312.84+312124.2	&	0.16	&	Flat	&	430	$\pm$	32	&	590	$\pm$	42	&	810	$\pm$	45	&	1400	$\pm$	78	&	4600	$\pm$	500	&	3400	$\pm$	370	\\
290	&	J033313.80+312005.3	&	1.31	&	0+I	&	0.86	$\pm$	0.06	&	1.8	$\pm$	0.10	&	3.2	$\pm$	0.17	&	7.1	$\pm$	0.37	&	68	$\pm$	7.1	&	57	$\pm$	18	\\
291	&	J033314.38+310710.9	&	2.72	&	0+I	&	0.23	$\pm$	0	&	1.3	$\pm$	0.09	&	2.5	$\pm$	0.19	&	5.5	$\pm$	0.39	&	130	$\pm$	14	&	830	$\pm$	120	\\
292	&	J033316.44+310652.5	&	1.56	&	0+I	&	$\cdots$			&	0.35	$\pm$	0.05	&	0.56	$\pm$	0.06	&	$\cdots$			&	23	$\pm$	2.4	&	1800	$\pm$	270	\\
293	&	J033316.65+310755.2	&	1.69	&	0+I	&	23	$\pm$	1.7	&	66	$\pm$	3.5	&	110	$\pm$	5.2	&	160	$\pm$	7.2	&	2000	$\pm$	210	&	6600	$\pm$	710	\\
294	&	J033317.85+310931.9	&	2.90	&	0+I	&	0.24	$\pm$	0.02	&	8.6	$\pm$	0.44	&	46	$\pm$	2.3	&	130	$\pm$	6.7	&	770	$\pm$	82	&	12000	$\pm$	1300	\\
295	&	J033320.32+310721.5	&	0.61	&	0+I	&	6.0	$\pm$	0.42	&	23	$\pm$	1.1	&	33	$\pm$	1.5	&	42	$\pm$	2.1	&	200	$\pm$	21	&	360	$\pm$	45	\\
296	&	J033327.29+310710.2	&	1.82	&	0+I	&	3.4	$\pm$	0.24	&	9.0	$\pm$	0.46	&	10	$\pm$	0.51	&	16	$\pm$	0.79	&	1800	$\pm$	200	&	5500	$\pm$	590	\\
297	&	J033330.41+311050.6	&	-1.04	&	II	&	1300	$\pm$	91	&	1300	$\pm$	92	&	1200	$\pm$	80	&	1200	$\pm$	81	&	980	$\pm$	100	&	520	$\pm$	59	\\
298	&	J033341.29+311341.0	&	-0.93	&	II	&	100	$\pm$	7.2	&	100	$\pm$	6.7	&	90	$\pm$	5.8	&	95	$\pm$	6.2	&	170	$\pm$	19	&	180	$\pm$	25	\\
299	&	J033346.92+305350.1	&	-1.17	&	II	&	4.9	$\pm$	0.34	&	4.1	$\pm$	0.31	&	3.7	$\pm$	0.31	&	3.9	$\pm$	0.31	&	6.2	$\pm$	0.67	&	$\cdots$			\\
300	&	J033351.07+311227.8	&	-2.16	&	III	&	42	$\pm$	3.0	&	29	$\pm$	2.4	&	19	$\pm$	2.1	&	13	$\pm$	2.1	&	5.5	$\pm$	0.60	&	$\cdots$			\\
301	&	J033401.66+311439.8	&	-1.16	&	II	&	100	$\pm$	7.1	&	95	$\pm$	6.8	&	100	$\pm$	6.7	&	120	$\pm$	7.5	&	110	$\pm$	11	&	91	$\pm$	16	\\
302	&	J033430.78+311324.4	&	-1.25	&	II	&	15	$\pm$	1.1	&	13	$\pm$	0.96	&	13	$\pm$	0.93	&	15	$\pm$	1.0	&	13	$\pm$	1.4	&	$\cdots$			\\
303	&	J033449.84+311550.3	&	-1.48	&	II	&	76	$\pm$	5.8	&	51	$\pm$	4.7	&	37	$\pm$	4.1	&	37	$\pm$	4.2	&	52	$\pm$	5.4	&	40	$\pm$	11	\\
304	&	J033915.81+312430.7	&	-1.07	&	II	&	22	$\pm$	1.5	&	19	$\pm$	1.3	&	14	$\pm$	1.1	&	19	$\pm$	1.3	&	34	$\pm$	3.6	&	$\cdots$			\\
305	&	J034001.49+311017.3	&	-1.54	&	II	&	15	$\pm$	1.1	&	12	$\pm$	0.9	&	9.0	$\pm$	0.79	&	8.4	$\pm$	0.8	&	7.5	$\pm$	0.82	&	$\cdots$			\\
306	&	J034109.13+314437.9	&	-0.89	&	II	&	670	$\pm$	49	&	570	$\pm$	48	&	540	$\pm$	39	&	520	$\pm$	39	&	930	$\pm$	98	&	2200	$\pm$	230	\\
307	&	J034114.11+315946.3	&	-2.09	&	III	&	50	$\pm$	3.5	&	32	$\pm$	3.0	&	23	$\pm$	2.7	&	19	$\pm$	2.7	&	9.6	$\pm$	1.0	&	$\cdots$			\\
308	&	J034119.19+320203.8	&	-1.45	&	II	&	23	$\pm$	1.7	&	21	$\pm$	1.5	&	17	$\pm$	1.3	&	16	$\pm$	1.3	&	10	$\pm$	1.1	&	$\cdots$			\\
309	&	J034124.42+315327.9	&	-1.06	&	II	&	4.0	$\pm$	0.28	&	3.9	$\pm$	0.26	&	3.5	$\pm$	0.25	&	3.7	$\pm$	0.27	&	3.0	$\pm$	0.42	&	$\cdots$			\\
310	&	J034139.17+313610.7	&	-1.15	&	II	&	51	$\pm$	3.5	&	36	$\pm$	3.1	&	30	$\pm$	2.9	&	53	$\pm$	3.6	&	150	$\pm$	16	&	$\cdots$			\\
311	&	J034141.09+314804.6	&	-0.16	&	Flat	&	4.7	$\pm$	0.33	&	6.6	$\pm$	0.39	&	8.0	$\pm$	0.45	&	11	$\pm$	0.55	&	13	$\pm$	1.4	&	$\cdots$			\\
312	&	J034153.26+315019.2	&	-0.38	&	II	&	4.0	$\pm$	0.29	&	4.3	$\pm$	0.28	&	4.7	$\pm$	0.29	&	6.8	$\pm$	0.38	&	12	$\pm$	1.3	&	$\cdots$			\\
313	&	J034155.71+314811.4	&	0.28	&	Flat	&	24	$\pm$	1.7	&	32	$\pm$	1.9	&	42	$\pm$	2.3	&	62	$\pm$	3.2	&	140	$\pm$	15	&	250	$\pm$	43	\\
314	&	J034157.44+314836.7	&	-1.20	&	II	&	580	$\pm$	48	&	530	$\pm$	39	&	480	$\pm$	34	&	480	$\pm$	34	&	330	$\pm$	34	&	180	$\pm$	29	\\
315	&	J034157.75+314800.8	&	-1.11	&	II	&	92	$\pm$	6.7	&	85	$\pm$	5.9	&	77	$\pm$	5.3	&	98	$\pm$	6.2	&	140	$\pm$	15	&	$\cdots$			\\
316	&	J034158.67+314821.4	&	0.47	&	0+I	&	2.9	$\pm$	0.2	&	4.4	$\pm$	0.26	&	6.2	$\pm$	0.34	&	10	$\pm$	0.5	&	43	$\pm$	4.4	&	$\cdots$			\\
317	&	J034201.01+314913.4	&	-1.42	&	II	&	7.8	$\pm$	0.55	&	6.4	$\pm$	0.48	&	5.4	$\pm$	0.45	&	5.4	$\pm$	0.45	&	5.3	$\pm$	0.61	&	$\cdots$			\\
318	&	J034202.17+314802.1	&	1.23	&	0+I	&	1.3	$\pm$	0.09	&	1.6	$\pm$	0.10	&	2.9	$\pm$	0.17	&	9.1	$\pm$	0.45	&	140	$\pm$	14	&	240	$\pm$	36	\\
319	&	J034204.34+314711.6	&	-1.02	&	II	&	19	$\pm$	1.4	&	18	$\pm$	1.3	&	17	$\pm$	1.2	&	22	$\pm$	1.4	&	25	$\pm$	2.6	&	$\cdots$			\\
320	&	J034210.69+314705.6	&	-1.10	&	II	&	6.6	$\pm$	0.46	&	6.2	$\pm$	0.4	&	6.1	$\pm$	0.38	&	7.4	$\pm$	0.45	&	5.0	$\pm$	0.62	&	$\cdots$			\\
321	&	J034219.27+314327.0	&	-0.88	&	II	&	84	$\pm$	5.8	&	74	$\pm$	5.2	&	64	$\pm$	4.6	&	79	$\pm$	5.2	&	140	$\pm$	15	&	110	$\pm$	25	\\
322	&	J034220.32+320531.0	&	-1.26	&	II	&	15	$\pm$	1.1	&	12	$\pm$	0.95	&	9.2	$\pm$	0.85	&	12	$\pm$	0.94	&	20	$\pm$	2.1	&	$\cdots$			\\
323	&	J034223.33+315742.7	&	-2.08	&	III	&	52	$\pm$	3.7	&	34	$\pm$	3.1	&	25	$\pm$	2.9	&	22	$\pm$	2.8	&	9.4	$\pm$	1.0	&	$\cdots$			\\
324	&	J034227.12+314432.9	&	-1.22	&	II	&	21	$\pm$	1.5	&	15	$\pm$	1.2	&	11	$\pm$	1.0	&	8.7	$\pm$	1.0	&	56	$\pm$	6.0	&	$\cdots$			\\
325	&	J034232.10+315249.5	&	-1.43	&	II	&	1.8	$\pm$	0.12	&	1.4	$\pm$	0.11	&	1.2	$\pm$	0.12	&	1.5	$\pm$	0.12	&	1.1	$\pm$	0.28	&	$\cdots$			\\
326	&	J034232.91+314220.6	&	-1.19	&	II	&	39	$\pm$	2.8	&	33	$\pm$	2.4	&	26	$\pm$	2.1	&	27	$\pm$	2.2	&	51	$\pm$	5.4	&	$\cdots$			\\
327	&	J034233.13+315214.7	&	-1.07	&	II	&	3.9	$\pm$	0.27	&	3.1	$\pm$	0.23	&	2.4	$\pm$	0.22	&	2.9	$\pm$	0.24	&	7.7	$\pm$	0.85	&	$\cdots$			\\
328	&	J034234.19+315101.0	&	-1.19	&	II	&	3.4	$\pm$	0.24	&	2.5	$\pm$	0.20	&	1.7	$\pm$	0.19	&	1.7	$\pm$	0.19	&	7.1	$\pm$	0.79	&	$\cdots$			\\
329	&	J034236.47+315517.6	&	-1.21	&	II	&	3.1	$\pm$	0.22	&	2.4	$\pm$	0.19	&	2.3	$\pm$	0.19	&	2.5	$\pm$	0.24	&	3.6	$\pm$	0.45	&	$\cdots$			\\
330	&	J034244.50+315958.7	&	-1.49	&	II	&	2.7	$\pm$	0.18	&	2.3	$\pm$	0.17	&	2.0	$\pm$	0.15	&	1.8	$\pm$	0.16	&	1.1	$\pm$	0.28	&	$\cdots$			\\
331	&	J034249.18+315011.2	&	-1.54	&	II	&	26	$\pm$	1.9	&	20	$\pm$	1.6	&	17	$\pm$	1.4	&	17	$\pm$	1.5	&	15	$\pm$	1.6	&	$\cdots$			\\
332	&	J034254.67+314345.3	&	-2.22	&	III	&	26	$\pm$	1.8	&	17	$\pm$	1.5	&	12	$\pm$	1.4	&	7.5	$\pm$	1.4	&	3.6	$\pm$	0.55	&	$\cdots$			\\
333	&	J034255.95+315842.0	&	-0.84	&	II	&	510	$\pm$	38	&	430	$\pm$	35	&	450	$\pm$	32	&	650	$\pm$	40	&	190	$\pm$	20	&	2400	$\pm$	270	\\
334	&	J034256.05+315644.8	&	0.06	&	Flat	&	4.0	$\pm$	0.29	&	6.2	$\pm$	0.36	&	9.2	$\pm$	0.48	&	18	$\pm$	0.87	&	13	$\pm$	1.3	&	$\cdots$			\\
335	&	J034301.94+314435.6	&	-1.15	&	II	&	3.9	$\pm$	0.27	&	3.6	$\pm$	0.25	&	3.5	$\pm$	0.25	&	4.0	$\pm$	0.27	&	3.4	$\pm$	0.49	&	$\cdots$			\\
336	&	J034306.77+314820.5	&	-2.40	&	III	&	25	$\pm$	1.8	&	17	$\pm$	1.4	&	11	$\pm$	1.3	&	7.1	$\pm$	1.2	&	2.0	$\pm$	0.32	&	$\cdots$			\\
338	&	J034321.47+314246.3	&	-1.50	&	II	&	6.9	$\pm$	0.48	&	5.5	$\pm$	0.42	&	4.6	$\pm$	0.38	&	5.2	$\pm$	0.41	&	3.3	$\pm$	0.45	&	$\cdots$			\\
339	&	J034322.22+314613.6	&	-1.15	&	II	&	36	$\pm$	2.5	&	33	$\pm$	2.3	&	27	$\pm$	2.0	&	31	$\pm$	2.1	&	38	$\pm$	4.0	&	$\cdots$			\\
346	&	J034344.48+314309.3	&	-0.50	&	II	&	130	$\pm$	11	&	230	$\pm$	13	&	240	$\pm$	14	&	330	$\pm$	17	&	550	$\pm$	59	&	470	$\pm$	53	\\
376	&	J034406.47+314325.1	&	-2.39	&	III	&	22	$\pm$	1.7	&	16	$\pm$	1.3	&	10	$\pm$	1.2	&	7.1	$\pm$	1.1	&	2.1	$\pm$	0.30	&	$\cdots$			\\
415	&	J034427.90+322718.9	&	-1.18	&	II	&	25	$\pm$	1.8	&	24	$\pm$	1.6	&	18	$\pm$	1.4	&	25	$\pm$	1.6	&	29	$\pm$	3.1	&	$\cdots$			\\
442	&	J034435.34+322837.2	&	-0.16	&	Flat	&	0.42	$\pm$	0.04	&	1.3	$\pm$	0.07	&	0.74	$\pm$	0.08	&	1.1	$\pm$	0.11	&	3.9	$\pm$	0.47	&	$\cdots$			\\
465	&	J034443.06+313733.7	&	-1.29	&	II	&	7.4	$\pm$	0.58	&	5.8	$\pm$	0.51	&	5.0	$\pm$	0.43	&	5.4	$\pm$	0.45	&	9.5	$\pm$	1.0	&	$\cdots$			\\
474	&	J034452.00+322625.4	&	-1.62	&	III	&	14	$\pm$	1.3	&	12	$\pm$	0.87	&	8.6	$\pm$	0.77	&	7.5	$\pm$	0.73	&	5.7	$\pm$	0.64	&	$\cdots$			\\
489	&	J034513.51+322434.8	&	-1.14	&	II	&	5.9	$\pm$	0.41	&	4.7	$\pm$	0.36	&	3.6	$\pm$	0.33	&	3.8	$\pm$	0.33	&	5.5	$\pm$	0.63	&	$\cdots$			\\
496	&	J034533.47+314555.3	&	-1.24	&	II	&	4.1	$\pm$	0.31	&	3.7	$\pm$	0.27	&	3.1	$\pm$	0.26	&	3.2	$\pm$	0.25	&	4.9	$\pm$	0.57	&	$\cdots$			\\
498	&	J034536.83+322557.0	&	-1.32	&	II	&	140	$\pm$	12	&	120	$\pm$	9.0	&	84	$\pm$	7.9	&	97	$\pm$	7.9	&	200	$\pm$	21	&	$\cdots$			\\
500	&	J034548.27+322412.0	&	-0.97	&	II	&	1300	$\pm$	110	&	1200	$\pm$	91	&	930	$\pm$	83	&	990	$\pm$	78	&	3000	$\pm$	320	&	5500	$\pm$	580	\\
501	&	J034558.25+322647.5	&	-1.39	&	II	&	8.9	$\pm$	0.73	&	7.8	$\pm$	0.66	&	7.0	$\pm$	0.58	&	6.4	$\pm$	0.55	&	4.9	$\pm$	0.58	&	$\cdots$			\\
502	&	J034657.38+324917.4	&	-1.36	&	II	&	7.1	$\pm$	0.50	&	5.9	$\pm$	0.45	&	4.7	$\pm$	0.41	&	4.3	$\pm$	0.4	&	5.7	$\pm$	0.64	&	$\cdots$			\\
503	&	J034658.51+324658.9	&	-1.54	&	II	&	8.3	$\pm$	0.58	&	6.1	$\pm$	0.50	&	5.2	$\pm$	0.47	&	5.5	$\pm$	0.48	&	4.3	$\pm$	0.51	&	$\cdots$			\\
504	&	J034705.43+324308.5	&	0.51	&	0+I	&	89	$\pm$	6.3	&	110	$\pm$	7.1	&	130	$\pm$	7.4	&	170	$\pm$	9.7	&	690	$\pm$	73	&	930	$\pm$	100	\\
505	&	J034741.58+325144.1	&	1.30	&	0+I	&	560	$\pm$	49	&	1100	$\pm$	65	&	1800	$\pm$	92	&	2200	$\pm$	130	&	3900	$\pm$	410	&	11000	$\pm$	1200	

\enddata

\end{deluxetable}

\clearpage
\begin{deluxetable}{lccccccccccccc}
\tabletypesize{\scriptsize}
%\rotate
\tablecolumns{14}
\tablecaption{Properties of the Regions \label{tab:class}}
\tablewidth{0pt}
\tablehead{
	\colhead{Region}  & 
	\colhead{Area\tablenotemark{a}}  & 
	\multicolumn{2}{c}{YSOs} &
	\colhead{Density\tablenotemark{c}} & 
	\multicolumn{2}{l}{Class 0+I} &
	\multicolumn{2}{l}{Flat} &
	\multicolumn{2}{l}{Class II} &
	\multicolumn{2}{l}{Class III} &
	\colhead{Ratio} \\
	\colhead{}  & 
	\colhead{(pc\textsuperscript{2})}  & 
	\colhead{N$_{YSO}$} &
	\colhead{Frac.\tablenotemark{b}} &
	\colhead{(pc\textsuperscript{-2})} &
	\colhead{N} &
	\colhead{N/N$_{YSO}$\tablenotemark{d}} &
	\colhead{N} &
	\colhead{N/N$_{YSO}$\tablenotemark{d}} &
	\colhead{N} &
	\colhead{N/N$_{YSO}$\tablenotemark{d}} &
	\colhead{N} &
	\colhead{N/N$_{YSO}$\tablenotemark{d}} &
	\colhead{II/(0+I)}
}

\startdata
All of Perseus	&	73.5	&	369	&	1.00	&	5.0	&	70	&	0.19	&	32	&	0.09	&	231	&	0.63	&	36	&	0.10	&	3.3	\\
IC 348	&	6.85	&	143	&	0.39	&	21	&	11	&	0.08	&	5	&	0.03	&	104	&	0.73	&	23	&	0.16	&	9.5	\\
NGC 1333	&	3.24	&	104	&	0.28	&	32	&	28	&	0.27	&	12	&	0.12	&	59	&	0.57	&	5	&	0.05	&	2.1	\\
RC	&	63.4	&	122	&	0.33	&	1.9	&	31	&	0.25	&	15	&	0.12	&	68	&	0.56	&	8	&	0.07	&	2.2	\\
B1	&	7.62	&	22	&	0.06	&	2.9	&	10	&	0.45	&	3	&	0.14	&	8	&	0.36	&	1	&	0.05	&	0.8	\\
L1455	&	7.62	&	11	&	0.03	&	1.4	&	5	&	0.45	&	2	&	0.18	&	4	&	0.36	&	0	&	0.00	&	0.8	\\
Eastern Perseus	&	31.0	&	193	&	0.52	&	6.2	&	15	&	0.08	&	9	&	0.05	&	140	&	0.73	&	29	&	0.15	&	9.3	\\
Western Perseus	&	42.5	&	176	&	0.48	&	4.1	&	55	&	0.31	&	23	&	0.13	&	91	&	0.52	&	7	&	0.04	&	1.7	

\enddata

\tablenotetext{a}{Assuming a distance of 250 pc.}
\tablenotetext{b}{Fraction of the total number of YSOs in Perseus that are in the region.}
\tablenotetext{c}{N$_{YSO}$ per Area.}
\tablenotetext{d}{Fraction of the total number of YSOs in the region that are of the above classification.}
\end{deluxetable}

\end{document}